\pdfoutput=1

\documentclass[conference,compsoc]{IEEEtran}
\usepackage[compress]{cite}
  
\usepackage{xcolor,soul,framed} 
\usepackage{latexsym}
\usepackage{hyperref}

\usepackage[utf8]{inputenc}

\usepackage{graphicx}
\usepackage{subfigure}
\usepackage{amsmath}
\usepackage{amssymb}
\usepackage{amsthm}
\usepackage{bbding}
\usepackage{tcolorbox} 
\usepackage{varwidth} 
\tcbuselibrary{breakable} 
\tcbuselibrary{skins} 
\NewTColorBox{theobox}{o m +o}{
    enhanced,
    colframe=gray!5!white,
    interior empty,
    coltitle=white,
    fonttitle=\bfseries,
    colbacktitle=gray,
    extras broken={frame empty,interior empty},
    borderline={0.5mm}{0mm}{gray},
    sharp corners=downhill,
    breakable=true,
    top=4mm,
    before skip=3.5mm,
    attach boxed title to top left={yshift=-3mm,xshift=5mm},
    boxed title style={boxrule=0pt,sharp corners=all},varwidth boxed title,
    IfNoValueTF={#1}{title=#2}{title=#2:~#1},
    IfNoValueTF={#3}{}{#3}
}

\usepackage{algorithm}
\usepackage{algorithmic}

\usepackage{booktabs}
\usepackage{multirow}
\usepackage{makecell}
\usepackage{CJKutf8}
\title{Provably Robust and Secure Steganography in Asymmetric Resource Scenario}
\author{
\IEEEauthorblockN{Minhao Bai$^1$, Jinshuai Yang$^1$, Kaiyi Pang$^1$, Xin Xu$^1$, Zhen Yang$^2$, Yongfeng Huang$^{1,3}$}
\IEEEauthorblockA{ $^1$Department of electronic engineering,
Tsinghua University\\
 $^2$School of Cyberspace Security, Beijing University of Posts and Telecommunications\\
 $^3$Zhongguancun Laboratory}
}

\begin{document}
\newtheorem{lemma}{Lemma} 
\newtheorem{proposition}{Proposition} 
\newtheorem{theorem}{Theorem} 
\newtheorem{definition}{Definition} 
\maketitle

\begin{abstract}
To circumvent the unbridled and ever-encroaching surveillance and censorship in cyberspace,  steganography has garnered attention for its ability to hide private information in innocent-looking carriers.
Current provably secure steganography approaches require a pair of encoder and decoder to hide and extract private messages, both of which must run the same model with the same input to obtain identical distributions. 
These requirements pose significant challenges to the practical implementation of steganography, including limited access to powerful hardware and the intolerance of any changes to the shared input.
To relax the limitation of hardware and solve the challenge of vulnerable shared input, a novel and practically significant scenario with asymmetric resource should be considered, where only the encoder is high-resource and accessible to powerful models while the decoder can only read the steganographic carriers without any other model's input.
This paper proposes a novel provably robust and secure steganography framework for the asymmetric resource setting. Specifically, the encoder uses various permutations of distribution to hide secret bits, while the decoder relies on a sampling function to extract the hidden bits by guessing the permutation used. Further, the sampling function only takes the steganographic carrier as input, which makes the decoder independent of model's input and model itself. 
A comprehensive assessment of applying our framework to generative models substantiates its effectiveness. Our implementation demonstrates robustness when transmitting over binary symmetric channels with errors. 
\end{abstract}

\section{Introduction}


There is a lasting anxiety surrounding freedom of speech and data security. 
In communications where the vast majority of data is in plaintext, encrypted data may raise suspicion of monitors \cite{mazurczyk2017information,zielinska2014trends,katzenbeisser2016information}. Furthermore, in scenarios with stringent control, only plaintext may be permitted for transmission, while the transmission of any unverified encrypted content may be prohibited. 
Steganography, as a class of techniques that disguise private messages as innocent-looking carriers, can help to protect the privacy in the plaintext channels to evade surveillance and censorship \cite{anderson1998limits,provos2003hide}. 


Steganography is a game that a sender attempts to send messages to a receiver under the monitoring of a censor.
The sender leverages a steganographic encoder to hide the messages into some steganographic carriers that look like normal carriers in the plaintext channel, and sends the carriers to receiver via the chosen channel. 
During this procedure, the censor tries to selectively block those who send suspicious carriers. 
If the steganographic carriers are successfully received, the receiver leverages the steganographic decoder to extract hidden messages from them. 
In this game, the goal of steganography is to escape from censor's monitoring by producing the steganographic carriers that are as innocent as possible.

With the rapid evolution of AIGC (Artificial Intelligence Generated Content), plaintext channels such as images, audios and texts are filled with artificially produced carriers by sophisticated generative models. Most of the generative models compute a high-quality explicit probability distribution of the carriers based on a given history or prompt, and then sample from the distribution to generate plaintext. 
Attracted by the potent ability of producing high-quality plaintext, practical steganography methods \cite{sallee2003model,fridrich2009steganography,cachin2005digital,lubacz2014principles,RNNstega,METEOR,DISCOP,witt2023perfectly,fridrich2007practical} often leverage the generative models to 
produce steganographic data under the control of the secret message, which is usually bits. Many methods \cite{METEOR,DISCOP,witt2023perfectly} are designed to achieve such control while maintaining the quality of generated data. When the steganographic carrier (stegotext) is computationally indistinguishable from the normally generated carrier (covertext), we say 
the corresponding steganography method is computationally secure \cite{hopper,METEOR}, which means a practical censor has a negligible advantage in detecting the stegotext. 
%

Previous theoretical work \cite{hopper} proves that secure steganography implies one-way functions, which suggests that researchers utilize pseudo-random generators (PRGs) to achieve security. 
In the paradigm of current secure and practical steganography methods \cite{DISCOP, METEOR}, the encoder runs the model to obtain a distribution, and uses PRG and secret messages to sample from that distribution.
In the decoding process, the decoder runs the same model to derive an identical distribution, then use PRG with the same key to recover the sampling process from the received stegotexts and extract the secret messages. 
Moreover, reproducing the identical distributions often requires running an identical model fed with identical inputs such as history and prompt.
However, reproducing an identical distribution without any distortion between the encoder and decoder is challenging in practice. The problem is the necessity of powerful hardware and prefix sharing. 

\textbf{Limitation of hardware:} the execution of a sophisticated model requires a powerful computing device. This requirement was not considered difficult prior to the evolution of AIGC, because at that time the most powerful models had few parameters and could be easily deployed on consumer-level devices, even if the quality of the generative data was poor. However, the more recently developed generative models have massive parameters to generate human-like data, which cannot be used in consumer-level devices. Requiring both the encoder and decoder to have access to powerful devices may be unreasonable or unrealistic, leading to a limitation in the application of current steganography methods.


\textbf{Challenge of sharing:} for clarity, we utilize the term ``prefix'' to represent input of the model. To generate suspectless stegotext, the encoder often uses the context of the channel as the prefix. Once the prefix is modified, messages can hardly be correctly extracted since the models will not produce the identical distribution. However, sharing the prefix before each transmission is challenging unless the encoder and decoder have control of the context in the channel, such as a private chat, 
in which case the cryptography has been deemed a sufficiently adequate choice. 
In public plaintext channels, such as online forums or social applications, the context is continuously changing, resulting in a potential disagreement between the encoder and decoder on the prefix. In such situation, sharing the correct prefix becomes a challenge for current steganography methods.

Our goal is to relax the limitation of hardware and completely solve the challenge of sharing. In practice, the sender who has the ability to produce steganographic data is often less restricted, while the receiver is often in a strict surveillance environment. Therefore, we consider a weakened setting, asymmetric resource scenario, where only the encoder is high-resource and accessible to generative models while the decoder can only read the stegotext. Current secure methods are not applicable to this weakened setting, so we decide to construct a new framework for it. To solve the challenge of sharing, rather than designing a complex method to share the correct prefix, we prefer constructing a decoder that does not rely on the prefix.

In this paper, we propose a novel provably robust and secure steganography framework for the asymmetric resource scenario. The encoder uses various permutations of distribution to hide secret bits, and the decoder relies on a sampling function that takes only the stegotext as input to extract the hidden bits by guessing the permutation that the encoder uses. The design of the sampling function makes the decoder independent of prefix and model and helps the decoder to distinguish the stegotext from the covertext. In our framework, stegotext is generated by sampling from a selected permutation of the model's distribution, while the output is computationally indistinguishable from the model's common output. Additionally, if a small part of stegotext is randomly substituted, the decoder can still correctly extract the hidden bits.

Our contributions can be summarized as follows:

\begin{enumerate}
    \item \textbf{Relaxed limitation of hardware.} The proposed steganography framework is provably secure and applicable within the asymmetric resource scenario: the encoder has access to the generative model while the decoder does not.
    \item \textbf{No need of sharing prefix.} Our framework does not rely on any prefix. The decoder is able to find the stegotext from the plaintext channel and correctly extract the secret messages only from the received stegotext.
    \item \textbf{Robustness against substitution.} To the best of our knowledge, this framework is the first practical steganography work that is robust against substitution attacks that randomly substitute a minor part of the stegotext.
\end{enumerate}

\section{Background and Related Work}
\subsection{Preliminaries}

\subsubsection{Channel}

In most situations, steganography is constructed over a plaintext channel $\mathcal{C}$ with an alphabet $A$, which is the set of symbols that could potentially appear. Formally speaking, we say that a Channel $\mathcal{C}$ is a pair of algorithms $(M,s)$, where $M$ is the generative model that estimates a probability distribution defined over the alphabet $A$ and $s$ is the sampler that samples from the resulting distribution. 

Since the language models (LM) have the ability to produce human-like text, we focus on text-based channels, where the model $M$ is a LM and the alphabet $A$ is the set of tokens (that is words, parts of words, punctuation marks, emojis and so on) used by the corresponding LM.
In such channels, $M$ takes the history and prompt as input, both of which we refer as the term “prefix” as before, and it outputs a distribution $\mathcal{D}$, which represents the probabilities of the next tokens. 
The sampler $s$ takes the distribution $\mathcal{D}$ as input and outputs a token $a$ from the alphabet $A$ based on the distribution $\mathcal{D}$. We write the above process as $a \stackrel{s}{\longleftarrow} \mathcal{D}$. For the normal text generation process, the distribution of sampler's output is identical to the distribution $\mathcal{D}$.

\subsubsection{Steganography}

The main goal of steganography is to substitute the channel's normal sampler $s$ with a steganographic sampler $s'$ that is controlled by secret messages $m \in \{0,1\}^*$.
In most cases, we expect the generated stegotext ($a \stackrel{s'}{\longleftarrow} \mathcal{D}$) to be indistinguishable from the covertext ($a \stackrel{s}{\longleftarrow} \mathcal{D}$). Meanwhile, in traditional steganography system, the whole computation process should be completely recovered by the decoder. Due to the sampler $s'$ is pre-shared, the decoder can rerun the generation process to recover the secret messages $m$ based on the stegotext. 

Formally, a traditional steganography system consists of three algorithms: \textbf{KeyGen}, \textbf{Encode}, and \textbf{Decode}.
\begin{itemize}
    \item \textbf{KeyGen}($1^\lambda$) generates a key $k$ that is shared between the encoder and decoder.
    \item \textbf{Encode}($k,M,z,m$) employs the model $M$ and the prefix $z$ to predict the distribution of the next symbol. The key $k$ and covertexts $m$ are used to determine the transmission of symbol $a$. This algorithm substitutes the original sampler of language models.
    \item \textbf{Decode}($k,M,z,a$) employs the model $M$ and the prefix $h$ to predict the distribution of the received symbol. The key $k$ is utilized to extract the covertexts from symbol $a$.
\end{itemize}

The pre-shared information includes private key $k$, prefix $z$, language model $M$. 

\subsubsection{Security}



In most of steganography works, researchers often aim at achieving computational security.

\begin{definition}[{Computational Security}]
A steganography system is computationally secure if, for all probabilistic polynomial-time adversaries $\mathcal{A}$, the advantage of $\mathcal{A}$ in distinguishing between the output of steganography sampler $\mathcal{O}_s$ and the output of random sampler $\mathcal{O}_r$ is negligible. Formally, if
\begin{align}
    \left|\mathbb{P}\left\{\mathcal{A}^{\mathcal{O}_s}\left(1^{\lambda}\right) = 1\right\} - \mathbb{P}\left\{\mathcal{A}^{\mathcal{O}_r}\left(1^{\lambda} \right) = 1 \right\}\right| <  \text{negl}(\lambda),
\end{align}
where $\text{negl}(\lambda)$ is the negligible function that correlates to the secure parameter $\lambda$, we say that the steganography system is computationally secure.
\end{definition}

 A function $negl(\lambda)$ is considered negligible if, for any constant $c > 0$, there exists a large integer $N$ such that for all $\lambda > N$,
    $negl(\lambda) < \frac{1}{\lambda^c}$.


\subsubsection{Correctness}
A steganography system must be correct, ensuring that the probability of decoding errors is negligible. 
\begin{definition}[Correctness]
    Correctness means that the probability of decoding failure is negligible. Formally, if for any message $m$ and any prefix $z$, 
\begin{align}
    &\mathbb{P}\left\{\textbf{Decode}(k,\textbf{Encode}(k,M,z,m))=m \right\} \geq 1- \text{negl}(\lambda),
\end{align}
    we say the steganography system is correct.
\end{definition}

\subsubsection{Robustness against Substitution}
We define the robustness against substitution as follows.
\begin{definition}[$e$-Robustness against substitution]
    The steganography system is $e$-robust against substitution if the probability of successfully decoding from an $n$-bit string $\{b_n\} = \textbf{Encode}(k,M,z,m)$ after transmitting through a binary symmetric channel (BSC) $\mathcal{C}_e$ with the error probability $e$ is more than $1-\text{negl}(n)$.
    \begin{align}
        \mathbb{P}\{\textbf{Decode}(k,M,z,\mathcal{C}_e(\{b_n\})) = m\} \geq 1-\text{negl}(n) .
    \end{align}
\end{definition}

\subsection{Related Work}
\subsubsection{Theoretical Steganography}

The origins of steganography can be traced back to Shannon's work  \cite{shannon1949communication} on secrecy systems. The formal definition of steganography scenario is first proposed by Simmons \cite{simmons1984prisoners}, which is known as ``Simmon's Prisoners' Problem''. Based on Simmons' foundational work, early theoretical steganography methods \cite{Cachin,mittelholzer1999information,wang2008perfectly,moulin2003information} focused on solving steganography problem in the aspect of information theory. They measured the statistical distance between stegotext and covertext by Kullback-Leibler Divergence, and defined the perfect security as this divergence is 0. However, the perfect security is difficult to achieve. Instead, researchers turned to building steganography methods that are computationally secure against the probabilistic polynomial-time adversaries. 

A large proportion of theoretical steganography methods \cite{hopper2002provably,von2004public,van2003efficient,hopper,berndt2017algorithm,anderson1996stretching,christ2024undetectable} aiming at computational security relies on rejection sampling. This sampling method queries the oracle multiple times until the desired output is obtained. We use the construction 1 in \cite{hopper} as an example. This construction requires a hash function such that the hash of the output word ends in 0 or 1 with exactly the same probability, no matter what the distribution of words is. When embedding the secret bit 0, we can let a black-box language model generate a word, and if the hash of the word ends in 0, we just send that word.  Otherwise, we have the model generate another word and send that word. In the end, the hash of the sent word has a probability of $\frac{3}{4}$ ending in 0. After several generations, more than half of the sent words should have a hash value ending in 0. Although such hash functions cannot exist in a practical scenario, this method provides a significant part of the spirit for our work.

\subsubsection{Practical Steganography}
Compared to theoretical work, mainstream of the practical work \cite{METEOR,DISCOP,witt2023perfectly,RNNstega,dai2019towards,ziegler2019neural,zhang2021provably,huang2023dna,zeng2024towards} relies on sampling from explicit distributions provided by generative models. They require that the sender and receiver share the same distribution by running the same model with the same randomness and prefix, which makes them not that practical. The critical process is that the decoder simulates the sampling process of encoder and computes the secret messages from the shared information. 

However, a significant part of the practical work \cite{HC1, HC2, HC3, AC, SAAC, fang2017} is not rigorously secure. Besides, \cite{zhang2021provably} and \cite{huang2023dna} require that the distribution can be divided into 2 groups with equal probability, but such requirements are difficult to meet in practice. \cite{ziegler2019neural} and \cite{dai2019towards} are deterministic algorithms with randomness reuse problem \cite{METEOR}. Practical works that satisfy the definition of computational security only include METEOR \cite{METEOR} and DISCOP \cite{DISCOP}. Although we believe that the minimum entropy coupling based steganography \cite{witt2023perfectly} is also computationally secure, they do not give a rigorous proof of security or correctness.    

We take the DISCOP \cite{DISCOP} as an example of secure and practical steganography, which constructs several differently organized distribution copies to sample from. By sharing the distribution copies and the randomness used in sampling process, the decoder can simulate the sampling procedure of encoding. If the received samples match a simulation result re-sampled from one of the copies, then that copy is used by the encoder. Therefore, the encoder lets the copies represent different secret bits and the decoder has a probability to extract them. 


\section{Method}
\subsection{Settings of Asymmetric Resource Scenario}
We give an informal situation to illustrate the asymmetric resource scenario. Suppose you need to save a prisoner by telling him the escape plan, but the prisoner is under surveillance and any direct communication is not allowed. So you can walk around the cells chatting with others, but the prisoner will hear your voice. The prisoner will learn the escape plan from a portion of the chat he hears, although he may not know the content of the chat before and after you walk through his cell.

Formally, the asymmetric resource scenario involves 3 parties: \textbf{sender}, \textbf{receiver}, and \textbf{censor}. All of them can read messages from the channel but only the sender can send messages through the channel. The sender has a powerful computing platform while the receiver cannot afford to run the model but still can execute simple computation.
Before the communication starts, the sender and receiver should share a secret key and choose a public PRG. Then the sender lets the steganography encoder run the model to generate a stegotext, and sends it through the channel. The censor is a probabilistic polynomial-time algorithm which will selectively block those who send suspicious messages. The receiver knows nothing about the stegotext,  so he runs the steganography decoder, which tries to find the stegotext from each message transmitted through the channel, and then extracts secret messages.

To avoid the censor's suspicion, the steganography encoder should produce computationally indistinguishable stegotext. To correctly extract hidden messages from the real stegotext, the steganography decoder should first distinguish between stegotext and covertext and then distinguish between stegotext that represents hidden bit 0 and stegotext that represents hidden bit 1. 
Here we re-define the steganography system that is applicable within the asymmetric resource scenario.
\begin{definition}[Steganography in asymmetric resource scenario]
    A steganography system that is applicable within the asymmetric resource scenario consists of 3 algorithms: 
    \textbf{KeyGen}, \textbf{Encode}, and \textbf{Decode}.
\begin{itemize}
    \item \textbf{KeyGen}($1^\lambda$) generates a key $k$ to be shared between the encoder and decoder.
    \item \textbf{Encode}($k,M,z,m$) employs the model $M$ and the prefix $z$ to generate stegotext guided by secret messages $m$.
    \item \textbf{Decode}($k,a$) extracts secret messages $m$ from the stegotext $a$.
\end{itemize}
\end{definition}

Security property of the steganography system in asymmetric resource scenario is the same as the symmetric resource scenario. But the correctness property has to be re-defined as follows
\begin{definition}[Correctness in asymmetric resource scenario]
    The steganography system is correct if (i) the probability of decoding any bit from a covertext is negligible and (ii) the probability of  incorrectly decoding from a stegotext is negligible.
    \begin{align}
        &\mathbb{P}\left\{ \textbf{Decode}(k,a) \neq \emptyset | a \in \text{covertext}\right\} \leq \text{negl}(\lambda), \\
        &\mathbb{P}\left\{\textbf{Decode}(k,\textbf{Encode}(k,M,z,m))=m\right\} \geq 1- \text{negl}(\lambda).
    \end{align}
\end{definition}
In the following section, we discuss using hypothesis testing techniques to construct a steganography decoder and using permutations to construct a steganography encoder.
\begin{figure*}[ht]
    \includegraphics[width=1\linewidth]{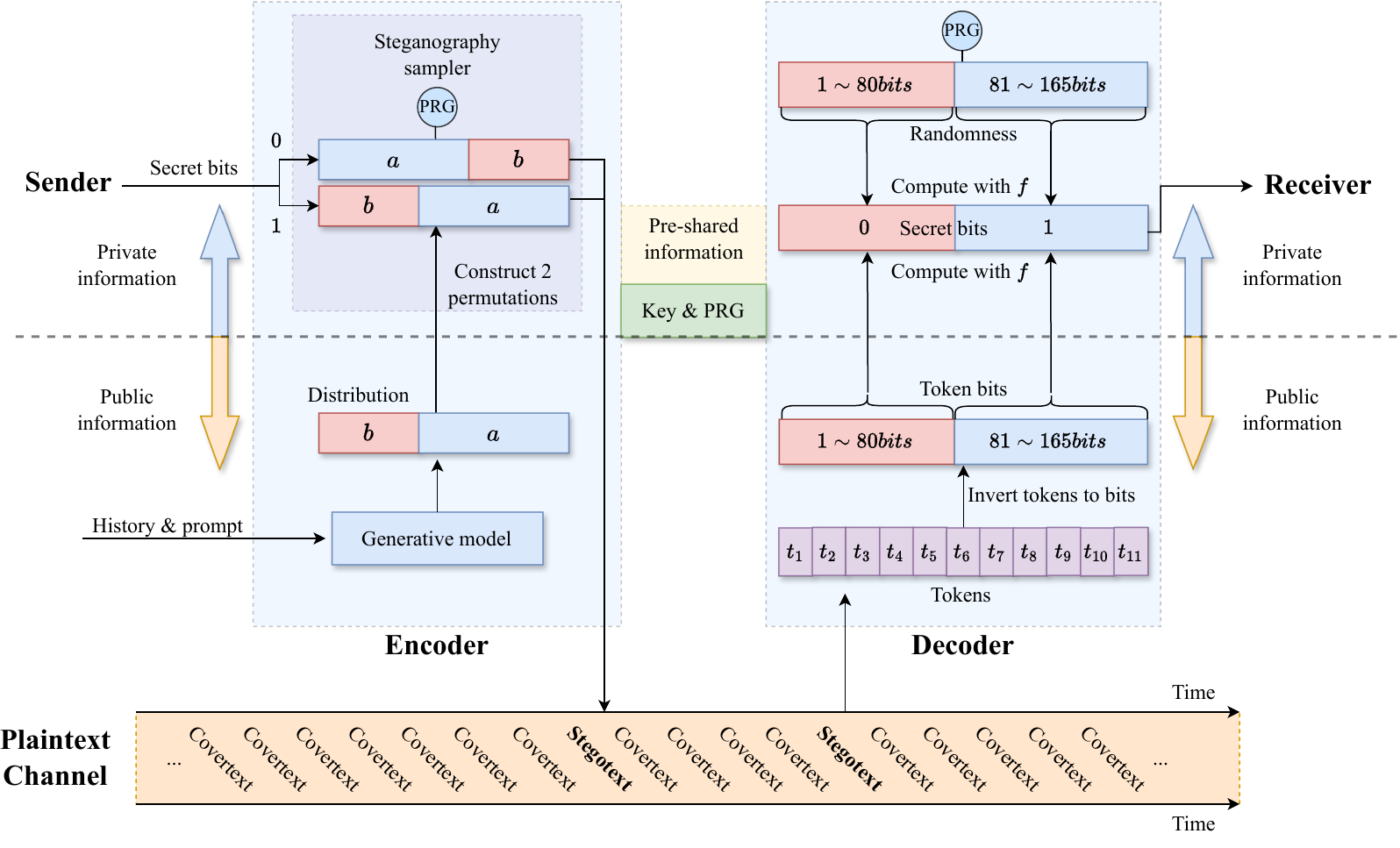}
    \caption{An overview of our steganography system.}
    \label{fig:非对称.drawio.pdf}
\end{figure*}
\subsection{Steganography in Asymmetric Resource Scenario}
\begin{figure}[t]
    \centering
    \includegraphics[scale = 0.6]{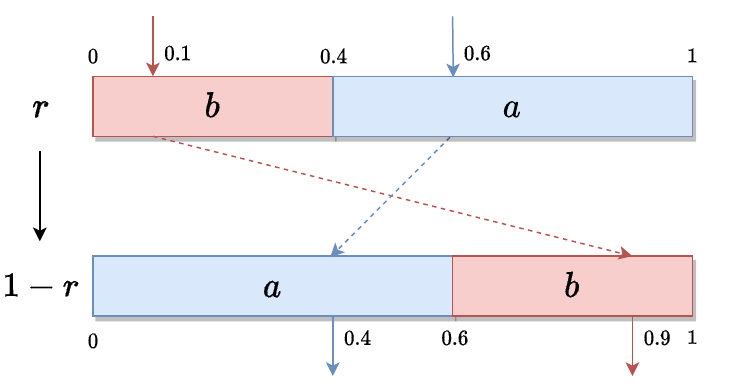}
    \caption{2 trivial permutations of the distribution of symbol $a$ and $b$.}
    \label{fig:顺序.drawio.pdf}
\end{figure}
\subsubsection{Construction}

We use 2 cascaded hypotheses to describe the situations that the decoder may encounter.
\begin{align}
    &\mathcal{H}_\emptyset: \text{There is no embedded bit.} \notag\\
    &\mathcal{H}_0: \text{There is an embedded bit 0.} \notag\\
    &\mathcal{H}_1: \text{There is an embedded bit 1.}\notag
\end{align}


For simplicity, we first decompose multi-variate distribution to multiple bi-variate distributions, and then consider the bi-variate distribution as shown in the Fig. \ref{fig:顺序.drawio.pdf}. Since such bi-variate distribution only has 2 different permutations denoted as $(a,b)$ and $(b,a)$ respectively, we let the permutation $(a,b)$ represent secret bit 0, and permutation $(b,a)$ represent secret bit 1. The construction of our steganography encoder is: (i) if the current secret bit is 0, the encoder samples from permutation $(a,b)$ to generate stegotext;
(ii) if the current secret bit is 1, the encoder samples from permutation $(a,b)$ to generate stegotext.

Since the decoder cannot use the same randomness to re-sample from the model, we try to substitute the re-sampling procedure by constructing a sampling function, which takes text and randomness as input and outputs a guess about the hidden bit with a confidence.
The decoder computes the sampling function from the shared randomness and should obtain a high confidence if it is a stegotext.

Here we construct an algorithm to distinguish between 3 hypotheses.
Let $f$ be a function that $\int_0^{x} f(r) - f(1-r) dr > 0$ for all $x \in (0,1)$. We define a sampling function $s$ as follows:
\begin{eqnarray}
    s = 
    \begin{cases}
        f(r) & \text{if $a$ is sampled,} \\
        f(1 - r) & \text{if $b$ is sampled.} \\
    \end{cases}
\end{eqnarray}
This function can be computed from the shared randomness $r$ and the received text. Since it inputs all of the information that decoder has, we hope to use its output to distinguish between the hypotheses.
We denote the expectation of sampling function $s$ computed from the stegotext that is generated by using permutation $(a,b)$ as $\mathbb{E}_r[s|\mathcal{H}_0]$, using permutation $(b,a)$ as $\mathbb{E}_r[s|\mathcal{H}_1]$ respectively. For the covertext that is not generated by steganography encoder, we denote the the expectation of sampling function $s$ as $\mathbb{E}_r[s|\mathcal{H}_\emptyset]$.
And we denote the probability of symbols $a$ and $b$ as $p(a)$ and $p(b)$. 

For a stegotext that is generated by using permutation $(a,b)$, the expectation of $s$ is 
\begin{align}
    &\mathbb{E}_{r}[s|\mathcal{H}_0] = \int_0^{p(a)} f(r) dr + \int_{p(a)}^1 f(1-r) dr.
\end{align}
For a covertext that is normally generated, the expectation of $s$ is
\begin{align}
    \mathbb{E}_r[s|\mathcal{H}_\emptyset] = \frac{1}{2}\left(\int_0^1 f(r) + f(1-r) dr\right).
\end{align}
There exists a gap between $\mathbb{E}_{r}[s|\mathcal{H}_0]$ and $\mathbb{E}_{r}[s|\mathcal{H}_\emptyset]$ as we require that $\int_0^{x} f(r) - f(1-r) dr > 0$ for all $x \in (0,1)$.
\begin{align}
    & \mathbb{E}_{r}[s|\mathcal{H}_0] - \mathbb{E}_{r}[s|\mathcal{H}_\emptyset]
    =  \int_0^{p(a)} f(r) - f(1-r) dr > 0.
\end{align}
Similarly, the gap between $\mathbb{E}_{r}[s|\mathcal{H}_1]$ and $\mathbb{E}_{r}[s|\mathcal{H}_\emptyset]$ is
\begin{align}
    & \mathbb{E}_{r}[s|\mathcal{H}_1] - \mathbb{E}_{r}[s|\mathcal{H}_\emptyset]
    =  \int_0^{p(a)} f(1-r) - f(r)dr < 0.
\end{align}
The decoder can compute sample mean $\Bar{s}$ from the received text and compare it with $\mathbb{E}_{r}[s|\mathcal{H}_\emptyset]$ to distinguish the 3 hypotheses. 

However, the sample mean $\Bar{s}$ can be close to $\mathbb{E}_{r}[s|\mathcal{H}_0]$ even if it is computed from a covertext, which will cause the decoder to accept $\mathcal{H}_0$ instead of $\mathcal{H}_\emptyset$. The probability of incorrectly rejecting $\mathcal{H}_\emptyset$ still exists. 
\begin{lemma}[{\bf Hoeffding's inequality}]\label{Hoeffding's inequality}
$X_1, X_2, \cdot\cdot\cdot , X_n$ are independent and identical random variables, and each $X_i$ is bounded by $[l,h], r \in [0,1]$.
Let $X = \frac{1}{n}\sum_{i=1}^n X_i$ and $\mu = \mathbb{E}[X_i]$, the probability that the sample mean $X$ deviates from the theoretical mean $\mathbb{E}[X]$ up to $t$ is 
\begin{align}
    \mathbb{P}\left(\left|X - \mu\right| \geq t\right) \leq 2\exp\left(-\frac{nt^2}{2\left(h-l\right)^2}\right).
\end{align}
\end{lemma}
According to Lemma \ref{Hoeffding's inequality}, the sample mean $\Bar{s}$ of a covertext has a negligible probability to be out of the interval $\left[\mathbb{E}_{r}[s|\mathcal{H}_\emptyset] - t, \mathbb{E}_{r}[s|\mathcal{H}_\emptyset] + t\right]$. Therefore, the decision bounds can be set as follows:

\begin{align}
    & \text{If }\exp\left(-\frac{n\left(\Bar{s} - \mathbb{E}_{r}[s|\mathcal{H}_\emptyset]\right)^2}{2(f_{max}(r)-f_{min}(r))^2}\right) \geq P_e , \text{ accept }\mathcal{H}_\emptyset;  \notag\\
    & \text{If } \bar{s} > \mathbb{E}_{r}[s|\mathcal{H}_\emptyset] \text{ and } \exp\left(-\frac{n\left(\Bar{s} - \mathbb{E}_{r}[s|\mathcal{H}_\emptyset]\right)^2}{2(f_{max}(r)-f_{min}(r))^2}\right)  < P_e , \notag\\
    &\text{ accept }\mathcal{H}_0; \notag\\
    & \text{If } \bar{s} < \mathbb{E}_{r}[s|\mathcal{H}_\emptyset] \text{ and } \exp\left(-\frac{n\left(\Bar{s} - \mathbb{E}_{r}[s|\mathcal{H}_\emptyset]\right)^2}{2(f_{max}(r)-f_{min}(r))^2}\right)  < P_e ,\notag\\
    &\text{ accept }\mathcal{H}_1; \notag
\end{align}

where $P_e$ is negligible and can be adjusted manually. These decision bounds ensure that the probability of falsely rejecting $\mathcal{H}_\emptyset$ is less than $P_e$. The probability of falsely accepting $\mathcal{H}_0$ and rejecting $\mathcal{H}_1$ is rather small, since the deviation between $\mathbb{E}_{r}[s|\mathcal{H}_0]$ and $\mathbb{E}_{r}[s|\mathcal{H}_1]$ is larger. The decoder is able to distinguish between 3 hypotheses with a probability of at least $1 - P_e$.  
\begin{algorithm}[t]
\caption{Main loop of steganography encoding}
    \renewcommand{\algorithmicrequire}{\textbf{Input:}}
    \renewcommand{\algorithmicensure}{\textbf{Output:}}
    \label{alg:encoding}
    \begin{algorithmic}[1]
    \REQUIRE Language model $M$, history $h$, alphabet $A$, pseudo-random generator $G_k$, secret bit $B$, error bound $P_e$, sampling function $f$.
    \ENSURE token list $T$.
    \STATE $P \longleftarrow M(h)$
    \STATE $seed, t_b, T \longleftarrow \emptyset$
    \STATE $s, n, next \longleftarrow 0$

    \REPEAT
        \STATE $P(t_b||0), P(t_b||1) \longleftarrow P$
        \STATE $r \longleftarrow G_k(\cdot)$
        \IF{ $r < \frac{(1-B)\cdot P(t_b||0)+B \cdot P(t_b||1)}{P(t_b||0) + P(t_b||1)}$}
        \STATE $next \longleftarrow B$
        \ELSE 
        \STATE $next \longleftarrow 1-B$
        \ENDIF
        \STATE $s \longleftarrow s || \left[ (1 - next)f(r) + nextf(1-r) \right]$
        \STATE $t_b \longleftarrow t_b || next$
        \STATE $n \longleftarrow n + 1$
        \IF{$n \text{ mod } \lceil \log_2(|A|) \rceil = 0$}
            \STATE $t \longleftarrow \text{Dec}(t_b)$
            \STATE $T_B \longleftarrow \emptyset$
            \STATE $T,h \longleftarrow T||t,h||t$
            \STATE $P \longleftarrow M(h)$
        \ENDIF
        \\
    \UNTIL{$\exp(-\frac{\text{sum}(s)^2}{8n}) \leq P_e \text{ }\AND\text{ } n \text{ mod} \lceil \log_2(|A|) \rceil = 0$}
    \RETURN $T$
    \end{algorithmic}
\end{algorithm}
\begin{algorithm}[t]
\caption{Main loop of steganography decoding}
    \renewcommand{\algorithmicrequire}{\textbf{Input:}}
    \renewcommand{\algorithmicensure}{\textbf{Output:}}
    \label{alg:encoding}
    \begin{algorithmic}[1]
    \REQUIRE alphabet $A$, pseudo-random generator $G_k$, error bound $P_e$, sampling function $f$, token list $T$.
    \ENSURE Secret bit $B$.
    \STATE $t_b \longleftarrow \text{Bin}(T)$
    \STATE $s, n \longleftarrow 0$

    \REPEAT
        \STATE $r \longleftarrow G_k(\cdot)$

        \STATE $s \longleftarrow s || \left[ (1 - t_b[n])f(r) + t_b[n]f(1-r) \right]$
        \STATE $n \longleftarrow n + 1$

    \UNTIL{$\exp(-\frac{\text{sum}(s)^2}{8n}) \leq P_e$}
    \IF{ $\text{sum}(s) > \mathbb{E}_{r}[s|\mathcal{H}_\emptyset]$}
        \STATE $B \longleftarrow 0$
    \ELSE
        \STATE $B \longleftarrow 1$
    \ENDIF
    \RETURN $B$
    \end{algorithmic}
\end{algorithm}

\subsubsection{Reduce the Multi-variate to Bi-variates}

Since the above discussion is based on bivariate distributions, it is necessary to find a method to decompose the multivariate distributions of LM into multiple bivariate distributions. 
A grouping method can be used. Constructing a complete binary tree whose leaves represent different tokens is an effective approach. At each node (not the leaves) in this tree, there are always two subtrees to choose from, reducing the multivariate distribution to multiple bivariate distributions.

One of the solutions is to group the tokens by the prefix of binary form of their IDs. For example, in \textsc{Llama2}'s tokenizer, the ID of ``Hello'' is $10,543$, and it can be converted to a binary form $010,1001,0010,1111$. The length is set as 15 bits since the vocabulary size of \textsc{Llama2}'s tokenizer is $32,000$, which is subtly smaller than $2^{15}$. Therefore, we can separate the tokens whose binary form begin with $0$ into the group 0 and those with $1$ into another group 1. In the group 0 we can further separate it into group 00 and group 01. This process will continue to run 15 loops to get a single token. That is to say, we convert a $32,000$-variate distribution into $15\times$ $2$-variate distributions.

If the decoder cannot even get access to the LM's tokenizer, we suggest converting the generated text into ASCII or Unicode. For example, token ``s'' is $0111,0011$ which starts with $0$ in ASCII. Each token can be translated into ASCII or Unicode like that. For the sake of clarity, the texts translated to binary form are called \textbf{token bits} and those secret messages in binary form are called \textbf{secret bits}. 


\subsubsection{Encoding \& Decoding}

If the current secret bit is $0$, a permutation $(\text{group 0}, \text{group 1})$ will be chosen as the distribution to sample from. Before the encoder confirms that this secret bit has been correctly embedded, it will continue to use the chosen permutation and accumulate $s$.

Once the sample mean $\bar{s}$ and the number of loops $n$ are high enough to satisfy the decision bound, it is confirmed that a bit has been embedded. Then, by comparing the sample mean $\bar{s}$ and the theoretical $\mathbb{E}[s|\mathcal{H}_\emptyset]$, the embedded bit can be derived by the decoder.

Details of the encoder are given in Alg. \ref{alg:encoding}, where $\text{Dec}(\cdot)$ is a function that converts a binary string to an integer. The encoder first gets a distribution from the language model, and then groups the vocabulary into group 0 and group 1. We denote the probability of sampling from group 0 as $P(0)$. The encoder gets a random number $r$ from the PRG and compares it with $\frac{(1-B)\cdot P(0)+B \cdot P(1)}{P(0) + P(1)}$ to sample a group. If we sample group 0 in the current step, we will sample from groups 00 and 01 in the next step. After several steps, a new token is sampled from the distribution, and the encoder must query the model with the new token appended to the history for the next distribution. The encoding procedure ends when the sum of $s$ satisfies $\exp(-\frac{sum(s)^2}{8n}) < P_e$. 

The decoding process is clearly included in the encoding process. The decoder only has to compute $s$ from the token bits and decide which hypothesis to accept at each step. This process only needs the token bits and the randomness generated by PRG and key. Even if the LM's tokenizer is forbidden, we can still convert the tokens to ASCII or Unicode. This way, the decoding is completely model free.

The overview of our steganography system is shown in Fig.\ref{fig:非对称.drawio.pdf}. The stegotext generated by the encoder is mixed into the covertext, and the decoder tries to extract hidden bits from each received text. For covertext, the decoder has a probability of at most $P_e$ to extract any bit. But for stegotext, the decoder can extract many hidden bits. 

\subsection{Properties of Our Framework}
\subsubsection{Capacity}\label{Capacity}

For the convenience, we use $p$ to denote the probability of sampling from group 0, and we denote the probability density function of $p$ as $\text{pdf}(p)$. Assuming we have 2 groups with probability $P$ and $1-P$, if we randomly choose one group as group 0, we have 
\begin{align}\label{18}
    \mathbb{P}\left\{ p = P\right\} = \mathbb{P}\left\{ p = 1 - P\right\}.
\end{align}
Therefore, the probability density function of $p$ is symmetric about $x = \frac{1}{2}$.
\begin{align}
    \text{pdf}(p) = \text{pdf}(1 - p).
\end{align}
We take into account the distribution of the model. In expectation, 
\begin{align}
    &\mathbb{E}_{p}\left[ \mathbb{E}_r\left[s|\mathcal{H}_0\right] - \mathbb{E}_r[s| \mathcal{H}_\emptyset] \right] \notag\\ 
    =& \int_0^1 \left(F(p) + F(1-p) - F(1) - F(0)\right) \text{pdf}(p) dp \notag\\
    =& \int_0^1 \left(F(p) + F(1-p)\right) \text{pdf}(p) dp - (F(1) + F(0)) \notag\\
    =& 2\int_0^1 F(p)\text{pdf}(p) dp - (F(1) + F(0)),
\end{align}
where the $F(\cdot)$ is the original function of $f(\cdot)$. 
According to the decision bounds, we have 
\begin{align}
    \exp\left( -\frac{n (\mathbb{E}_{p}\left[ \mathbb{E}_r\left[s|\mathcal{H}_0\right] - \mathbb{E}_r[s| \mathcal{H}_\emptyset] \right])^2}{2\left(f_{max}(r)-f_{min}(r)\right)^2}\right) < P_e.
\end{align}
Simplifying the above inequality we have
\begin{align}
    n > \frac{2\left(f_{max}(r)-f_{min}(r)\right)^2(-\ln(P_e))}{\left(2\int_0^1 F(p)\text{pdf}(p) dp - (F(1) + F(0))\right)^2},
\end{align}
where $\ln(\cdot)$ represents the natural logarithm function.
This inequality shows the expectation of steps we need to embed a single bit.
\subsubsection{Security}

The security of our framework is based on the pseudo-random generator. The critical point is whether the output of encoding algorithm is computationally indistinguishable from the original output of language model. We prove the security by a sequence of games and we focus on the distribution of their outputs.
\begin{theorem}
    The output of our framework is computationally indistinguishable from the normal output of the used model.
\end{theorem}
\textit{Proof.} $G_0 = G_1$: $G_0$ is the steganographic sampling procedure controlled by the secret bit, while $G_1$ is independent of secret bit. $P(T_B||0)$ and $P(T_B||1)$ are the probabilities of 2 groups. So the probability of sampling from the group $T_B||0$ in $G_0$ is 
\begin{align}
    & \mathbb{P}\{\text{sampling from group } T_B||0\} \notag\\
    = & \mathbb{P}\left\{r < \frac{P(T_B||0)}{P(T_B||0) + P(T_B||1)}\right\} \mathbb{P}\{B=0\} + \notag \\
    & \mathbb{P}\left\{r \geq \frac{P(T_B||1)}{P(T_B||0) + P(T_B||1)}\right\} \mathbb{P}\{B=1\}
    \notag \\
    = & \frac{P(T_B||0)}{P(T_B||0) + P(T_B||1)} \left(\mathbb{P}\{B=0\} + \mathbb{P}\{B=1\}\right) \notag \\
    = & \frac{P(T_B||0)}{P(T_B||0) + P(T_B||1)},
\end{align}
which is identical to the probability of sampling from the group $T_B||0$ in $G_1$. Therefore, the distribution of the output of $G_0$ and $G_1$ is the same. 
\begin{figure}[htbp]
\begin{theobox}{$G_0$:steganographic sampling}
\begin{algorithmic}[1]
    \STATE $P(t_b||0), P(t_b||1) \longleftarrow M(h)$
        \STATE $r \longleftarrow G_k(\cdot)$
        \IF{ $r < \frac{(1-B)\cdot P(t_b||0)+B \cdot P(t_b||1)}{P(t_b||0) + P(t_b||1)}$}
        \STATE $t_b \longleftarrow t_b||B$
        \ELSE 
        \STATE $t_b \longleftarrow t_b||(1-B)$
        \ENDIF
        \IF{$\text{len}(t_b) = \lceil \log_2(|A|) \rceil$}
            \RETURN $\text{Dec}{(t_b)}$
            \ENDIF
\end{algorithmic}
\end{theobox}
\begin{theobox}{$G_1$: removing secret bits}

\begin{algorithmic}[1]
    \STATE $P(t_b||0), P(t_b||1) \longleftarrow M(h)$
        \STATE $r \longleftarrow G_k(\cdot)$
        \IF{ $r < \frac{P(t_b||0)}{P(t_b||0) + P(t_b||1)}$}
        \STATE $t_b \longleftarrow t_b||0$
        \ELSE 
        \STATE $t_b \longleftarrow t_b||1$
        \ENDIF
        \IF{$\text{len}(t_b) = \lceil \log_2(|A|) \rceil$}
            \RETURN $\text{Dec}{(t_b)}$
            \ENDIF
\end{algorithmic}
\end{theobox}
\begin{theobox}{$G_2$: {from pseudo random to random.}}
\begin{algorithmic}[1]
    \STATE $P(t_b||0), P(t_b||1) \longleftarrow M(h)$
        \STATE $r \longleftarrow R(\cdot)$
        \IF{ $r < \frac{P(t_b||0)}{P(t_b||0) + P(t_b||1)}$}
        \STATE $t_b \longleftarrow t_b||0$
        \ELSE 
        \STATE $t_b \longleftarrow t_b||1$
        \ENDIF
        \IF{$\text{len}(t_b) = \lceil \log_2(|A|) \rceil$}
            \RETURN $\text{Dec}{(t_b)}$
            \ENDIF
\end{algorithmic}
\end{theobox}
\begin{theobox}{$G_3$: from multiple bi-variate to multi-variate.}
\begin{algorithmic}[1]
\STATE $P \longleftarrow M(h)$
        \STATE $r \longleftarrow R(\cdot)$
        \IF{ $r \in \text{interval of token }t$}
            \RETURN $t$
        \ENDIF
\end{algorithmic}
\end{theobox}
\begin{theobox}{{$G_4$: model random sampling.}}
\begin{algorithmic}[1]
        \STATE $P \longleftarrow M(h)$
        \STATE $\text{randomly sample token $t$ from $P$}$
        \RETURN $t$
\end{algorithmic}
\end{theobox}
\caption{Games used in the proof.}
\end{figure}

$G_1 \approx_c G_2$: $G_2$ uses random generator $R(\cdot)$ to substitute the pseudo-random generator $G_k(\cdot)$ in $G_1$. From the definition of pseudo-random generator, for any probabilistic polynomial time (p.p.t.) adversary $\mathcal{A}$, 
\begin{align}
    \left|\mathbb{P}\left\{ \mathcal{A}^{G_k(\cdot)}(1^k) = 1\right\} - \mathbb{P}\left\{ \mathcal{A}^{R(\cdot)}(1^k) = 1\right\}\right| \leq \text{negl}(k).
\end{align}
Assuming the existence of a p.p.t. adversary $\mathcal{A}'$, which can distinguish the output of $G_1$ and $G_2$ with a noticeable probability, we can use the adversary $\mathcal{A}'$ to construct another p.p.t. adversary $\mathcal{A}''$ to distinguish between the output of random generator $R(\cdot)$ and pseudo-random generator $G_k(\cdot)$. $\mathcal{A}''$ runs $G_1$ and $G_2$ with random generator $R(\cdot)$ and pseudo-random generator $G_k(\cdot)$. Though now  $\mathcal{A}''$ can neither distinguish between $R(\cdot)$ and $G_k(\cdot)$ nor $G_1$ and $G_2$, he can query $\mathcal{A}'$ for each pair of output of $G_1$ and $G_2$. Since $\mathcal{A}'$ can distinguish the output of $G_1$ and $G_2$ with a noticeable probability, the adversary $\mathcal{A}''$ can distinguish the output of random generator $R(\cdot)$ and pseudo-random generator $G_k(\cdot)$ with a noticeable probability, which contradicts the definition of pseudo-random generator. Therefore, such adversary $\mathcal{A}'$ does not exist, and the distribution of output of $G_1$ and $G_2$ is computationally indistinguishable.

$G_2 = G_3$: $G_2$ needs to sample several times to output a single token, while $G_3$ directly samples from the model's output distribution. Assuming that a token $t$ has the binary form $\{b_1, b_2, \cdots, b_{\lceil \log_2(|A|) \rceil}\}$, the probability of sampling this token in $G_2$ is 
\begin{align}
& \mathbb{P} \left\{ \text{output token } t \right\} = \prod_{i=1}^{\lceil \log_2(|A|) \rceil} \mathbb{P}\{\text{sample from group } b_{1:i} \} \notag\\
 =& \frac{P(b_{1:1})}{1} \times \frac{P(b_{1:2})}{P(b_{1:1})} \times \cdots \times \frac{P(b_{1:\lceil \log_2(|A|) \rceil})}{P(b_{1:(\lceil \log_2(|A|) \rceil-1)})} \notag \\
 =& P(b_{1:\lceil \log_2(|A|) \rceil}) = P(t) = \mathbb{P}\{r \in \text{interval of $t$}\},
\end{align}
which is identical to the probability of sampling the token $t$ directly from the output distribution of the model in $G_3$. Therefore, distributions of the outputs of $G_2$ and $G_3$ are identical.

$G_3 = G_4$: $G_3$ uses a specific sampling method, while $G_4$ does not specify the sampling method. As the sampling output obeys the distribution of model, and the probability of random number $r$ falls into the interval of token $t$ is just the probability of token $t$ in the distribution of model, the distributions of outputs of $G_3$ and $G_4$ are identical.

Based on the above, we have proved that $G_0 = G_1 \approx_c G_2 = G_3 = G_4$. So the output of our steganography system is computationally indistinguishable from the original model's output. Therefore, our system is computationally secure. \qed

\subsubsection{Correctness} \label{Error Mechanism}
The decoding failures can be classified into 2 types: decoding a secret bit from covertext, or decoding a wrong secret bit from stegotext.
\begin{theorem} \label{corr1}
    In this steganography framework, the probability of decoding a secret bit  from a covertext is negligible. 
\end{theorem}
\textit{Proof.} $P_e$ represents the probability of decoding a secret bit from a covertext, and it is set to a negligible value, which has proved the above theorem. \qed
\begin{theorem}\label{corr2}
    In this steganography framework, the probability of decoding a secret bit 0 from a stegotext that contains secret bit 1 is negligible. 
\end{theorem}
\textit{Proof.}
Assuming that the encoder tries to embed a secret bit $0$ in $n$ token bits,
the sample mean $\bar{s}$ should be larger than $\mathbb{E}[s|\mathcal{H}_\emptyset]$ and satisfy that $\exp\left(-\frac{n\left(\bar{s} - \mathbb{E}[s|\mathcal{H}_\emptyset]\right)^2}{2(f_{max}(r)-f_{min}(r))^2}\right) < P_e$. We want to compute the probability of  $\bar{s} < \mathbb{E}[s|\mathcal{H}_\emptyset]$ under the condition of $\exp\left(-\frac{n\left(\bar{s} - \mathbb{E}[s|\mathcal{H}_\emptyset]\right)^2}{2(f_{max}(r)-f_{min}(r))^2}\right) < P_e$, in which case the decoder will extract a wrong bit.


From Hoeffding's inequality, the probability of $\bar{s} < \mathbb{E}[s|\mathcal{H}_0] - t$ is at most
\begin{align}
    \mathbb{P}\left\{ \bar{s} < \mathbb{E}[s|\mathcal{H}_0] - t\right\} \leq \exp\left(-\frac{nt^2}{f(0)-f(1)}\right).
\end{align}
If $\bar{s}$ is less than $ \mathbb{E}[s|\mathcal{H}_\emptyset]$, then $t$ should be larger than $\mathbb{E}[s|\mathcal{H}_\emptyset] - \mathbb{E}[s|\mathcal{H}_0]$. Therefore, the probability of $\bar{s} < \mathbb{E}[s|\mathcal{H}_\emptyset]$ is negligible in $n$. Since $P_e$ is negligible, $n$ should be large enough to make $\mathbb{P}\left\{ \bar{s} < \mathbb{E}[s|\mathcal{H}_0] - t\right\}$ negligible. Finally, the probability of decoding a secret bit 0 from a stegotext that contains secret bit 1 is negligible.  \qed

From Thm. \ref{corr1} and \ref{corr2}, the probability of decoding failure is negligible, proving that our framework is correct.

\subsubsection{Robustness against Substitution}

We discuss the ``substitution attack'' that keeps the length of messages. For example, the encoder sends a stegotext ``The cat sat on the map'' but the decoder receives ``The cat sat on the cap''. In the case of letter-level substitution, the decoder receives a message of the same length but some token bits are flipped. 
To discuss the influence of substitution formally, we consider a BSC $\mathcal{C}_e$ that has a probability $e$ to flip a text bit. For simplicity, we denote a constant $K = 2 \int_0^1 F(p)\text{pdf}(p) dp - F(0) - F(1)$, where $F(\cdot)$ is the original function of $f(\cdot)$ and $\text{pdf}(p)$ is the probability density function of $p$.

\begin{theorem}
    Our steganography framework is at least $(\frac{(1 - \frac{\sqrt{2}}{2})K}{f_{max}(r)-f_{min}(r)} -\frac{1}{n})$-robust against substitution.
\end{theorem}
\textit{Proof.} When transmitting a $n$-bit string $\{b_n\}$ through a BSC $\mathcal{C}_e$, the expectation of flipped token bits is
\begin{align}
    \mathbb{E}[d_H(\mathcal{C}_e(\{b_n\}),\{b_n\})] = ne,
\end{align}
where the $d_H(\cdot)$ represents the Hamming distance.

According to Hoeffding's inequality, the probability of flipping more than $ne+1$ token bits or more is at most
\begin{align}
    & \mathbb{P}[d_H(\mathcal{C}_e(\{b_n\}),\{b_n\}) \geq ne + 1]
     \leq 2\exp(-2n),
\end{align}
Therefore, if $n$ is sufficiently large, the probability of flipping more than $ne+1$ token bits is negligible.
Assuming that there are $ne+1$ flipped token bits in a $n$ bit string corresponding to a secret bit, when the sampling function $s$ is computed on these flipped token bits, the value of $s$ is originally $f(r_i)$ but now $f(1-r_i)$, or originally $f(1 - r_i)$ but now $f(r_i)$. 
We denote the sample mean of the function computing from $\{b_n\}$ as $\bar{s}$, and that computing from $\mathcal{C}_e(\{b_n\})$ as $\bar{s}'$. Then we denote the index set of error bits as $E$. From the above derivation, if $ne+1$ bits are flipped, we have 
\begin{align}
   \left|\bar{s} - \bar{s}'\right| = &\left| \sum_{i \in E} \left[f(r_i)-f(1-r_i)\right](1-2b_i) \right| \notag\\
    \leq & \left| \sum_{i \in E} \left[f(r_i)-f(1-r_i)\right] \right|\left|(1-2b_i)\right| \notag\\
    \leq & (ne+1) (f_{max}(r)-f_{min}(r)).
\end{align}

Comparing the expectation of $s$ under the condition of $\mathcal{H}_0$ and a zero-error channel $\mathcal{C}_0$ with the expectation of $s$ under the condition of $\mathcal{H}_\emptyset$ and an error channel $\mathcal{C}_e$, we have 
\begin{align}
    & \mathbb{E}_r[s|\mathcal{H}_0,\mathcal{C}_e] - \mathbb{E}_r[s|\mathcal{H}_\emptyset] \notag\\
    \geq & \mathbb{E}_r[s|\mathcal{H}_0,\mathcal{C}_0] - \frac{1+ne}{n} (f(0)-f(1)) - \mathbb{E}_r[s|\mathcal{H}_\emptyset] \notag\\
    = & F(p) + F(1-p) - F(0) - F(1) - \notag \\ 
    & \frac{1+ne}{n} (f_{max}(r)-f_{min}(r)).
\end{align}
Then we take expectation on $p$,
\begin{align}
   &\mathbb{E}_p\left[ \mathbb{E}_r[s|\mathcal{H}_0,\mathcal{C}_e] - \mathbb{E}_r[|\mathcal{H}_\emptyset]\right] \notag\\
   \geq &2 \int_0^1 F(p)\text{pdf}(p) dp - F(0) - F(1) - \notag \\ 
   & \frac{1+ne}{n} (f_{max}(r)-f_{min}(r)).
\end{align}
Combining this inequality with the decision bounds, we have
\begin{align}\label{new bound}
    & \exp\left( -\frac{n\left( \mathbb{E}_p\left[ \mathbb{E}_r[s|\mathcal{H}_0,\mathcal{C}_e] - \mathbb{E}_r[s|\mathcal{H}_\emptyset]\right]\right)^2}{2(f_{max}(r)-f_{min}(r))^2} \right) \notag\\
    \leq & \exp\left( -\frac{n\left(K - \frac{1+ne}{n} (f_{max}(r)-f_{min}(r))\right)^2}{2(f_{max}(r)-f_{min}(r))^2}\right) 
\end{align}
From this inequality we find that if  
\begin{align}\label{40}
    K - \frac{1+ne}{n} (f_{max}(r)-f_{min}(r)) \geq \frac{\sqrt{2}K}{2},
\end{align}
the $\exp$ term is negligible in $n$, which requires that 
\begin{align}
    e \leq \frac{(1 - \frac{\sqrt{2}}{2})K}{f_{max}(r)-f_{min}(r)} -\frac{1}{n}.
\end{align}
Therefore, our system is at least $(\frac{(1 - \frac{\sqrt{2}}{2})K}{f_{max}(r)-f_{min}(r)} -\frac{1}{n})$-robust against substitution. \qed

In practice, the decoder may not know whether the stegotext has been modified. There are several ways to remind the decoder about the state of the stegotext, such as using an error detection code. However, we prefer to use the internal robustness of our framework to correctly decode from the modified stegotext. 
\begin{figure*}[ht]
    \centering
    \subfigure[Temperature $=1.00$]{
    \centering
        \includegraphics[width=0.315\textwidth]{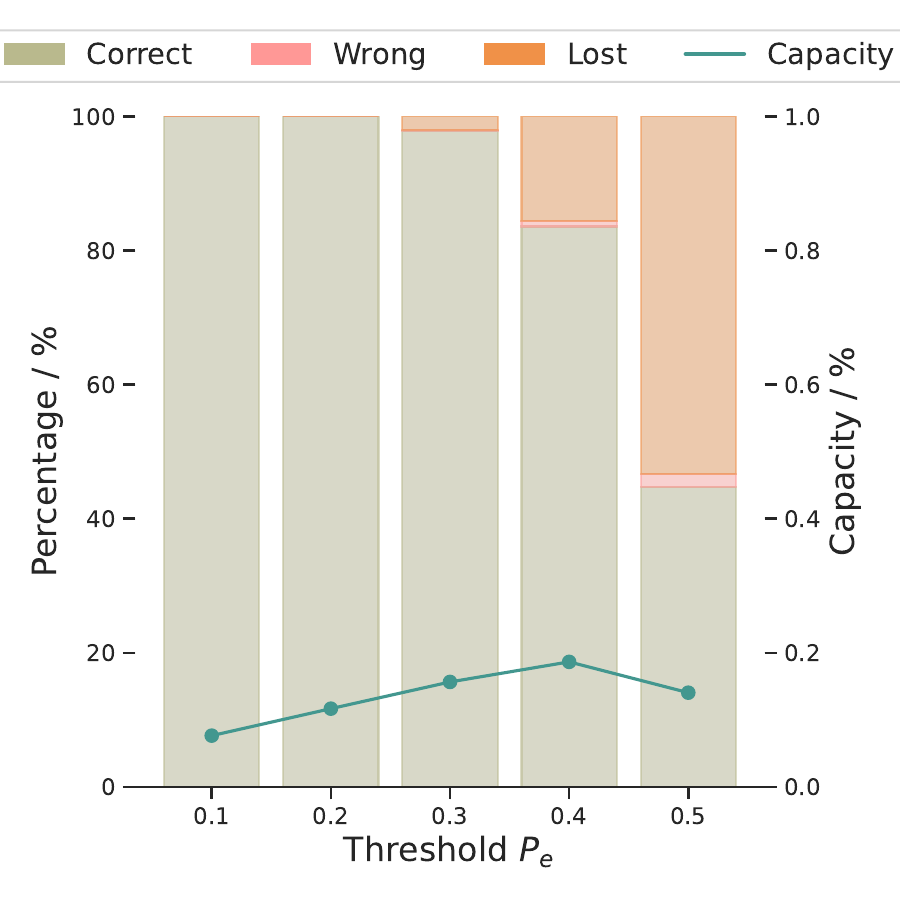}
        \label{}
    }
    \hfill
    \subfigure[Temperature $=1.05$]{
    \centering
        \includegraphics[width=0.315\textwidth]{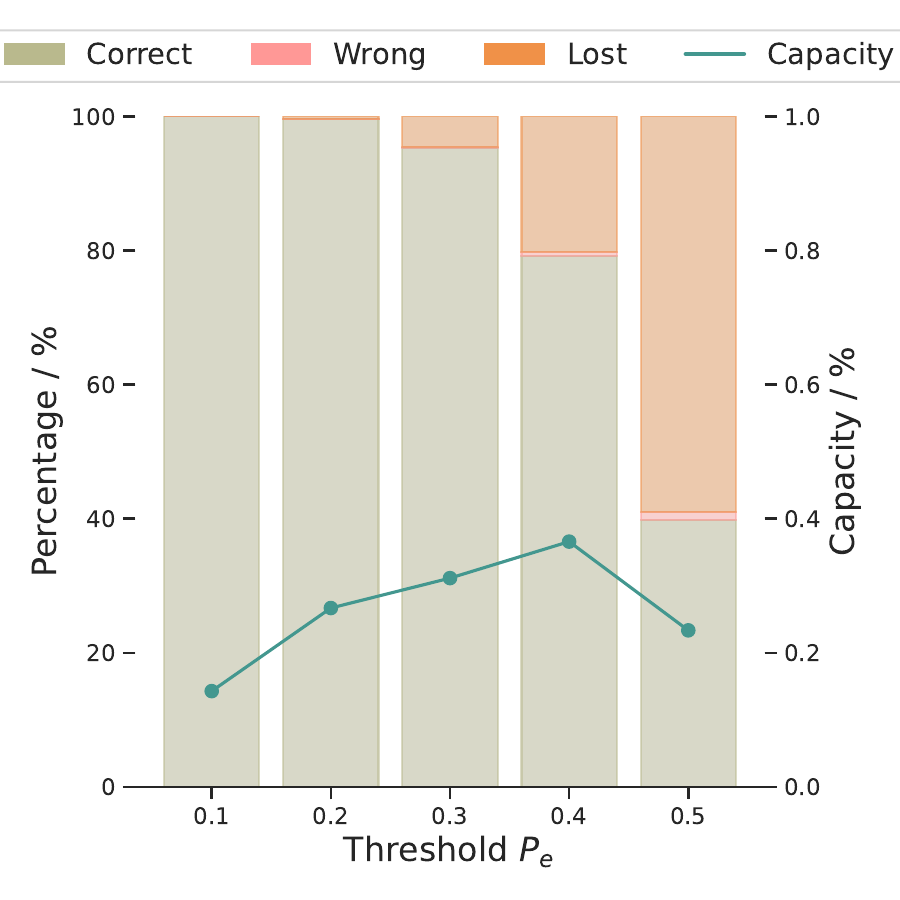}
        \label{}
    }
    \hfill
    \subfigure[Temperature $=1.10$]{
    \centering
        \includegraphics[width=0.315\textwidth]{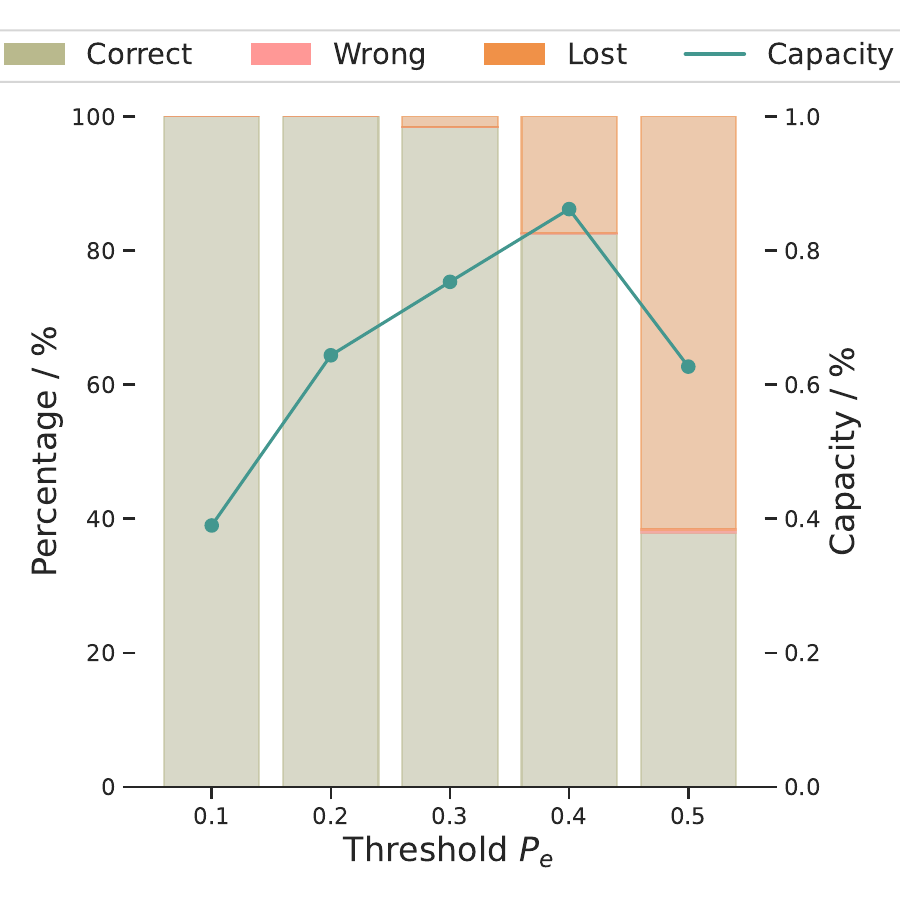}
        \label{}
    }
    \caption{Relationship between error bound $P_e$ ($0.1 \sim 0.5$) and bit error rate. }
    \label{fig:ber}
\end{figure*}
\begin{figure}[ht]
    \centering
    \includegraphics[width=\linewidth]{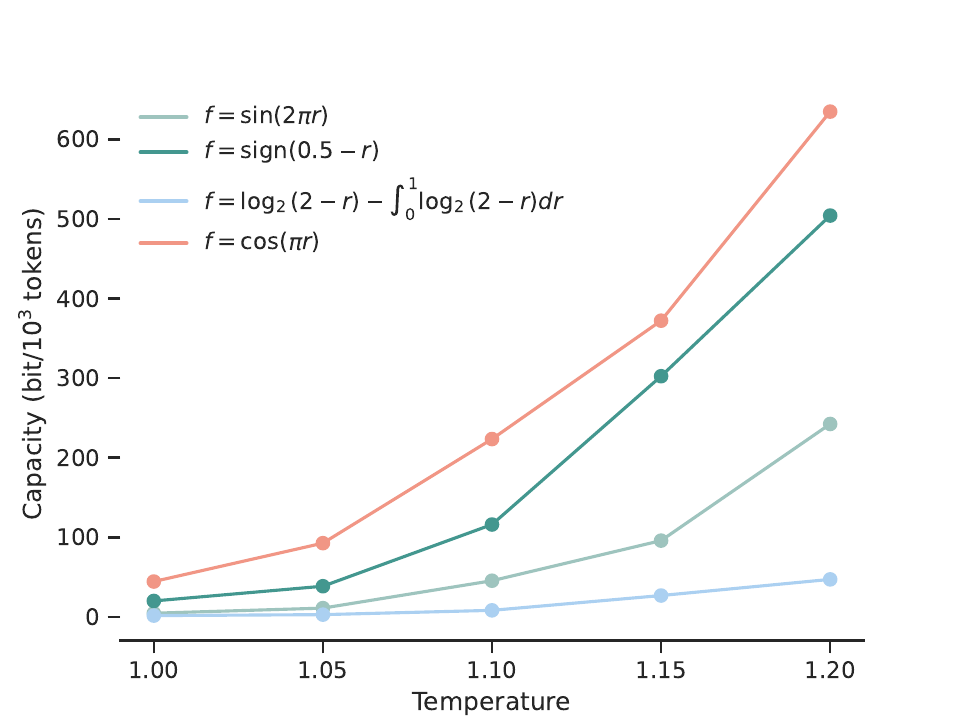}
    \caption{Capacity of different functions. }
    \label{fig:capacity-functions}
\end{figure}
The encoder, as a party that can send text into the channel and read text from the channel, can estimate the error probability and try to decrease the error bound $P_e$ to survive in transmission. According to inequality \ref{new bound}, the encoder can change the decision bounds
to make sure that the decoder can decode correctly. Or the encoder simply uses lower $P_e$. From inequality \ref{new bound} and \ref{40}, we have
\begin{align}
    &\exp\left( -\frac{n\left(K - \frac{1+ne}{n} (f(0)-f(1))\right)^2}{2(f(0)-f(1))^2}\right) \notag\\
    <& \exp\left( -\frac{\frac{nK^2}{2}}{2(f(0)-f(1))^2}\right) \notag\\
    <& P_e^{\frac{1}{2}}.
\end{align}
Therefore, the encoder can use $P_e^2$ as new error bound to guarantee that the decoder correctly decodes from the modified stegotext.

\section{Experiment}

\subsection{Settings}

\textbf{Metrics.} As a covert communication system, we measure the embedding capacity and encoding/decoding rate. Since our framework generates innocent-looking texts, we measure the perplexity (PPL) and entropy of the texts to verify that whether the statistics of the output change or not. We use the model that generates these output to compute the perplexity and entropy. The perplexity reflects the fluency of text, lower is better. The entropy represents the diversity of generated texts.

\textbf{Models.} We utilize 6 large language models (LLMs): \textsc{Llama2}-7B \cite{llama2}, \textsc{Mistral}-7B \cite{Mistral}, \textsc{Qwen2}-7B \cite{qwen2}, \textsc{Glm4}-9B \cite{glm4}, \textsc{Llama3}-8B \cite{llama3} and \textsc{Mpt}-7B \cite{mpt} as our stegotext generators. Limited by the memory of our GPUs, the models with more than 10 billion parameters can not be deployed on our devices.

\textbf{Configs.} We test our framework under 5 different temperature parameters ($1.0 \sim 1.2$) in order to figure out the relationship between entropy and embedding capacity. The prompt we used is \textit{Write 100 totally different tweet-like long comments.} This prompt is loaded into a instruction template and then becomes the input of LLMs.
Each data point represents the mean value of $10,000$ samples. 

\textbf{Hardwares.} We run all of the experiments on a server with 4 $\times$ NVIDIA A5000 GPUs (32GB RAM) and 24 $\times$ Intel Xeon w5-3423 CPUs.

\subsection{Preparations}

To find a suitable function $f$ and the error bound $P_e$ for our steganography system, we first test several basic functions and then test the relationship between the theoretical error bound and the BER in practice.

\begin{table*}[ht]
    \centering
    \caption{Overall Performance of Our Steganography System when Applied to Language Models}
    \begin{tabular}{c|c|c|c|c|c|c|c}
    \toprule[1.5pt]
    \midrule
        Model & Vocabulary size & Temperature & \makecell{Embedding capacity \\ (bit/$10^3$ token)} & \makecell{Encoding rate \\ (s/$10^3$ token)} & \makecell{Decoding rate \\ (s/$10^3$token)} & Perplexity & \makecell{Entropy \\ (bit/token)}\\ \midrule
        \multirow{5}{*}{\textsc{Llama2}} & \multirow{5}{*}{32000} &
            1.00 & 1.7142 & 69.5517& 0.1458 & 1.7072 & 0.6003 \\
        & & 1.05 & 3.2393 & 69.4213 & 0.1557 & 1.8207 & 0.7092\\
        & & 1.10 & 11.6175 & 69.5814 & 0.1465 & 1.9710 & 0.8497\\
        & & 1.15 & 68.5556 & 69.8412 & 0.1554 & 2.2591 & 1.0422\\
        & & 1.20 & 274.2626 & 69.2419 & 0.1478 & 3.7711 & 1.7247\\
        \midrule
        \multirow{5}{*}{\textsc{Llama3}} & \multirow{5}{*}{128000} &
         1.00 & 4.3932	& 77.1481 & 0.1611 & 2.0535 & 0.8540\\
        & & 1.05 & 11.9744 & 77.2511 & 0.1495 & 2.2781 & 1.0337 \\
        & & 1.10 & 50.9060 & 77.4706 & 0.1503 & 2.8414 & 1.3225\\
        & & 1.15 & 170.8803 & 77.7482 & 0.1666 & 3.5912 & 1.6596\\
        & & 1.20 & 462.6667 & 77.9605 & 0.1543 & 15.1243 & 3.8362\\
        \midrule
        \multirow{5}{*}{\textsc{Mistral}} & \multirow{5}{*}{32000} &
            1.00 & 6.3134 & 72.5419 & 0.1573 & 2.1374 & 0.9361\\
        & & 1.05 & 12.6068 & 72.4871 & 0.1544 & 2.3522 & 1.1226 \\
        & & 1.10 & 35.4009 & 72.2528 & 0.1499 & 2.9496 & 1.4410 \\
        & & 1.15 & 108.3932 & 72.5147 & 0.1510 & 5.4760 & 2.1837 \\
        & & 1.20 & 376.5874 & 72.9758 & 0.1506 & 22.5325 & 4.2987\\
        \midrule
        \multirow{5}{*}{\textsc{Glm4}} & \multirow{5}{*}{151329} &
         1.00 & 35.4037 & 90.3119 & 0.1639 & 2.3301 & 1.8867\\
        & & 1.05 & 90.8120 & 90.0081 & 0.1592 & 2.7715 & 2.3906\\
        & & 1.10 & 253.2821 & 90.6484 & 0.1591 & 3.6328 & 3.1738\\
        & & 1.15 & 469.5726 & 90.2186 & 0.1497 & 5.2812 & 4.3281 \\
        & & 1.20 & 652.7265 & 90.7294 & 0.1619 & 9.9606 & 6.2070\\
        \midrule
        \multirow{5}{*}{\textsc{Mpt}} & \multirow{5}{*}{50254} &
        1.00 & 6.0684 & 88.8987 & 0.1620 & 2.4888 & 1.0869\\
        & & 1.05 & 16.3419 & 88.3576 & 0.1647 & 2.8499 & 1.2930 \\
        & & 1.10 & 53.4872 & 88.7728 & 0.1502 & 3.4980 & 1.6895 \\
        & & 1.15 & 177.5128 & 88.1508 & 0.1481 & 6.2031 & 2.4648 \\
        & & 1.20 & 360.0940 & 88.1427 & 0.1552 & 20.6406 & 4.1523 \\
        \midrule
        \multirow{5}{*}{\textsc{Qwen2}} & \multirow{5}{*}{151643} &
         1.00 & 44.4036 & 71.4669 & 0.1558 & 2.2506 & 2.0778 \\
        & & 1.05 & 92.6325 & 72.9514 & 0.1610 & 2.5745 & 2.4659 \\
        & & 1.10 & 223.3675 & 72.4473 & 0.1697 & 2.9258 & 2.9008 \\
        & & 1.15 & 372.0769 & 71.2423 & 0.1583 & 4.1113 & 3.7476 \\
        & & 1.20 & 634.7094 & 72.4674 & 0.1664 & 5.8098 & 4.8946 \\
        \midrule
        \bottomrule[1.5pt]
    \end{tabular}
    
    \label{tab:my_label}
\end{table*}
\begin{table}[htb]
    \centering
    \caption{The Metrics of Covertexts Generated with the Same Configuration.}
    \begin{tabular}{c|c|c|c|c}
    \toprule[1.5pt]
    \midrule
        Model & Temperature & Perplexity & \makecell{Entropy \\ (bit/token)} & \makecell{Generation rate \\ ($10^3$ token/s)}
        \\ 
        \midrule
        \multirow{5}{*}{\textsc{Llama2}} 
        & 1.00 & 1.7253 & 0.6200 & 69.5404\\
        & 1.05 & 1.7894 & 0.6896 & 69.5521\\
        & 1.10 & 2.0066 & 0.8504 & 69.5746\\
        & 1.15 & 2.2774	& 1.0473 & 69.5408\\
        & 1.20 & 3.7490 & 1.6699 & 69.6191\\
        \midrule
        \multirow{5}{*}{\textsc{Llama3}} 
        & 1.00 & 2.0840 & 0.8863 & 76.8092\\
        & 1.05 & 2.2953 & 1.0198 & 76.6723\\
        & 1.10 & 2.8152 & 1.3193 & 76.6330\\
        & 1.15 & 3.6021 & 1.6885 & 76.5169\\
        & 1.20 & 15.2117 & 3.8302 & 77.1174\\
        \midrule
        \multirow{5}{*}{\textsc{Mistral}} 
        & 1.00 & 2.1494 & 0.9039 & 72.4695\\
        & 1.05 & 2.3508 & 1.1281 & 72.2360	\\
        & 1.10 & 3.1013 & 1.5237 & 72.5594	\\
        & 1.15 & 5.4175 & 2.3048 & 72.4958	\\
        & 1.20 & 22.5539 & 4.3922 & 72.4463\\
        \midrule
        \multirow{5}{*}{\textsc{Glm4}} 
        & 1.00 & 2.3301 & 1.8672 & 89.4464	\\
        & 1.05 & 2.7539 & 2.3516 & 89.4465\\
        & 1.10 & 3.4863 & 3.0703 & 89.6007\\
        & 1.15 & 5.2117 & 4.3211 & 89.3666\\
        & 1.20 & 10.3359 & 6.3398 & 89.3154\\
        \midrule
        \multirow{5}{*}{\textsc{Mpt}} 
        & 1.00 & 2.5046 & 1.0639 & 87.7857\\
        & 1.05 & 2.8713 & 1.2764 & 87.5573\\
        & 1.10 & 3.4727 & 1.5790 & 87.7506\\
        & 1.15 & 6.1898 & 2.3581 & 87.6666\\
        & 1.20 & 20.2656 & 4.0829 & 87.6691\\
        \midrule
        \multirow{5}{*}{\textsc{Qwen2}} 
        & 1.00 & 2.2553 & 2.0619 & 71.9145\\
        & 1.05 & 2.4993 & 2.3837 & 72.2306\\
        & 1.10 & 3.1252 & 2.9624 & 72.0331\\
        & 1.15 & 4.0391 & 3.8375 & 72.0565\\
        & 1.20 & 5.9162 & 4.9008 & 71.9757\\
        \midrule
        \bottomrule[1.5pt]
    \end{tabular}
    
    \label{tab:base}
\end{table}
\subsubsection{On the Sampling Function $f$}
We choose some representative functions that are integrable over the interval $[0,1]$, such as $\sin(2 \pi x)$, $\text{sign}(0.5 -r)$, $\log_2(2-r)-\int_0^1\log_2(2-r)dr$, and $\cos(\pi x)$. We control these functions such that their integral over the interval $[0, 1]$ is $0$ to maintain a certain degree of comparability. All of the stegotexts are generated by \textsc{Qwen2}. 

The results are shown in Fig. \ref{fig:capacity-functions}. 
$\sin(2 \pi r)$ shows the lowest capacity, since its value approaches $0$ when the $r$ approaches $1$. As we discussed in Section \ref{intuition}, the extreme values of random number $r$ have a more significant effect on our decision. However, $\sin(2 \pi r)$ represses this profit, leading to a low capacity. As a piece of additional evidence, applying $\sin(2 \pi r)$ into the optimization problem under the assumption that $p$ is uniformly distributed, it produces the same value as $\cos(\pi r)$.

Logarithm function $\log_2(2-r)-\int_0^1\log_2(2-r)dr$ also produces a relatively low capacity, possibly because its value is not $0$ when the random number $r$ is close to $0.5$. This may cause some bias when applying to LLMs' distribution. 
To our surprise, the sign function $\text{sign}(0.5 -r)$ produces a comparable capacity. Theoretically, this function should have the best performance when applying to a balanced distribution, but the LLMs' distributions are far from balanced. Because this function is not consistent at $0.5$, we suppose that is the reason for a loss of capacity. 
$\cos(\pi r)$ is a suitable function for LLMs' distributions. It produces the largest capacity, because it assigns values near $0$ to unimportant random numbers near $0.5$, and assigns high (or low) values to random numbers near $0$ (or $1$). 

\subsubsection{On the Error Bound $P_e$}
We then test the relationship between the theoretical error bound and the BER in practice to obtain a reasonable $P_e$. When the error bound $P_e$ is set extremely high, the decoder has a probability of extracting several wrong bits and losing the rest of the embedded bits. As shown in Fig. \ref{}, the wrong bits are only a small percentage. Even if the error bound $P_e$ is set to $0.5$, the wrong bits are still less than $2\%$. In all cases, we can find at most $4$ bit errors in a single transmission. So, if the error bound cannot be set to a negligible value or a negligible function of the secure parameter due to some implementation constraints, there is still a method to save the bit error rate: erase the last few bits. The encoder and decoder can try several times to decode some messages from innocent covertexts and see at most how many ``secret'' bits it can decode. So this number of bits could be a consensus between encoder and decoder. When the decoder tries to decode a stegotext, it has to extract the same number of bits from the end of the extracted bits. However, this process can reduce the capacity, which leads to a low communication rate.

\subsection{Results of the Proposed Framework}

\subsubsection{On the Capacity \& Entropy}
Results are shown in Tab.\ref{tab:my_label}. We observed a significant gain in embedding capacity when increasing the temperature parameter. The increased temperature brings higher entropy and makes the distributions more balanced. 
In most cases, the probability is concentrated on $1 \sim 2$ tokens. For a large fraction of the encoding steps, the encoder samples from an extreme distribution such as $p(a) = 0.99$, $p(b) = 0.01$, which has low entropy. No matter what permutation is used, the contribution of such steps to the sample mean $\bar{s}$ is really small.

However, the problem of low entropy may come from the inefficiency of the tokenizer. If the entropy is extremely low, there must be some strong relationships between the last token and the current token, which means that they should have been merged into a single token. Meanwhile, we find that the entropy of human twitter texts is $2.9514$ bits (measured by \textsc{Llama2}), which is much higher than the \textsc{Llama2}'s output. It seems that the LLMs are usually outputting some uninformative sentences, which is different from human's behaviour. When applying our framework to those LLMs whose vocabulary size is larger, such as \textsc{Qwen2} and \textsc{Glm4}, our framework often gets better performance and the generated text is more similar to twitter.


\subsubsection{On the Encoding \& Decoding Rate}
   
The encoding rate is mainly limited by the LLMs' generation speed. During encoding and decoding, the speed of processing tokens is stable. Since the decoding procedure gets rid of LLMs' complex forward computation, the decoding rate is much higher than the encoding rate. Although our encoding algorithm requires to loop 15 times or more during the generation of a single token, it does not impose significant burdens. On average, our framework needs less than 1 ms to process each token. 

Our implementation is based on Python codes, so the performance can still be largely optimized. If using C/C++ codes to substitute the steganography sampler part of the Python codes, the encoding rate will be more similar to the generation rate of the model. 
Our implementation is based on Python code, so the performance can still be optimized to a large extent. By using C/C++ code to replace the steganography sampler part of the Python code, the encoding rate will be more similar to the generation rate of the model. 

\subsubsection{On the Linguistic Quality of Stegotexts}

Comparing the Tab.\ref{tab:base} and Tab.\ref{tab:my_label} we find that the linguistic quality of stegotexts is similar to the innocent covertexts directly generated by the LLMs. This phenomenon supports our proof of security, which indicates that the difference between the stegotexts and the covertexts is negligible. 

\begin{table}[th]
    \centering
    \caption{Robustness against Random Substitution}
    \begin{tabular}{c|c|c|c}
    \toprule[1.5pt]
    \midrule
        \makecell{Error probability $e$ \\ (\%)} & \makecell{Correct Ratio \\ (\%)}& \makecell{Wrong Ratio \\ (\%)}& \makecell{Lost Ratio \\ (\%)}  \\
        \midrule
         0 & 99.9 & 0.0 & 0.0 \\
         1 & 99.9 & 0.0 & 0.0 \\
         2 & 99.9 & 0.0 & 0.0 \\
         3 & 99.9 & 0.0 & 0.0 \\
         4 & 99.9 & 0.0 & 0.0 \\
         5 & 99.9 & 0.0 & 0.0 \\
         6 & 99.9 & 0.0 & 0.0 \\
         7 & 99.9 & 0.0 & 0.0 \\
         8 & 97.5 & 0.1 & 2.4 \\
         9 & 92.6 & 0.0 & 7.4 \\
         10 & 91.7 & 0.0 & 8.3 \\
    \midrule
    \bottomrule[1.5pt]
    \end{tabular}
    
    \label{tab:robust}
\end{table}

\begin{table*}[ht]
    \centering
    \caption{Comparison with Baselines}
    \begin{tabular}{l|ccccccc}
    \toprule[1.5pt]
    \midrule
         Method \& Hyperparameter& Perplexity & \makecell{Entropy \\ (bit/token)}  & \makecell{Embedding capacity \\ (bit/$10^3$ token)} & \makecell{Encoding rate \\ (s/$10^3$ token)} & \makecell{Decoding rate \\ (s/$10^3$ token)} & Robustness\\
         \midrule
         DISCOP \cite{DISCOP}, non-recursive version& 2.4425 & 2.2471 & 968.9187 & 92.7573 & 92.1107 & \XSolidBrush\\
         DISCOP \cite{DISCOP}, recursive version& 2.3198 & 2.1107 & 1752.7986 & 8927.1584 & 8717.2002 & \XSolidBrush\\
         METEOR \cite{METEOR}, without reorder& 2.3954 & 2.2208 & 847.8570 & 89.7093 & 89.8521 & \XSolidBrush\\
         METEOR \cite{METEOR}, with reorder& 2.4518 & 2.3575 & 1158.3141 & 618.9911 & 619.7428 & \XSolidBrush\\
         \midrule
         Red-Green List \cite{kirchenbauer2023watermark}, $\gamma=0.25, \delta=2$ & 3.2461 & 2.8121 & - & 90.0011 & 0.1778 & \CheckmarkBold \\
         Undetectable Watermark \cite{christ2024undetectable}, $\lambda = 16$ & 2.2443 & 1.9344 & - & 90.6256 & 0.1577 & \CheckmarkBold\\
         \midrule
         Ours, BER $< 10^{-5}$ & 2.2506 & 2.0778 & 35.4037 & 90.3119 & 0.1639 & \CheckmarkBold\\
    \midrule
    \bottomrule[1.5pt]
    \end{tabular}
    
    \label{compare}
\end{table*}

\subsubsection{On the Robustness against Substitution}

In this part of the experiment, we consider the case where the sender knows that the stegotext has been modified, since he can read and send messages through the plaintext channel. In this case, only the encoder needs to reduce the error bound to decode correctly, and no other change in the encoding or decoding scheme is necessary. 

Since in the inequality \ref{new bound} and \ref{40}, computing $$K = 2 \int_0^1 F(p)\text{pdf}(p) dp - F(0) - F(1)$$ requires knowledge of the distribution of LLMs. From the above preparations, most of the density is concentrated in the vicinity of 1, and with some numerical calculations the value of $K$ is estimated to be $0.69$ in \textsc{Qwen2}. According to these statistics, our steganography system is at least $4.9\%$-robust, which is enough for some occasional bit flips during transmission. 
Therefore, we tested the robustness on channels with error probability $0 \sim 0.1$. All of the stegotext are generated by \textsc{Qwen2} with temperature $1.0$. The encoder set the error bound $P_e = 0.01$ while the decoder set the error bound $P_e = 0.1$. 

The results are displayed in Tab. \ref{tab:robust}. Our steganography system is stable for channels with error probability $0.1 \sim 0.7$, and still has a correct rate of more than 90\% even when 10\% of the token bits are reversed. In all situations, the wrong bits are extremely small, we observed only $1 \sim 2$ wrong bits in each experiment. The correct rate has an explicit decrease when the error probability increases from $7\%$ to $8\%$. Results show that the robust limit of our system is around $7\%$.

\subsection{Comparison with Existing Methods}
In this section, we compare the performance of our steganography system with current provably secure steganography methods. In addition, we compare with some of the watermarking methods since their application scenario is similar to ours. We use \textsc{Qwen2} as the generative model and the temperature is set to $1.0$.

METEOR \cite{METEOR} and DISCOP \cite{DISCOP} are used as the steganography baselines. Red-Green List \cite{kirchenbauer2023watermark} and Undetectable Watermark \cite{christ2024undetectable} are used as our watermark baselines. We follow the suggestions in the codes of Red-Green List to set the parameters $\gamma=0.25, \delta=2$, where $\gamma$ denotes the ratio of green list and $\delta$ denotes the negative impact on the text quality. As for Undetectable Watermark, we set $\lambda = 16$ in order to produce a relatively short watermarked output, where $\lambda$ is the secure parameter. Results are shown in Tab. \ref{compare}. 

For steganography methods, since they require the encoder and decoder to share the distributions, their capacity is much higher than ours. However, they may not take full advantage of this shared distribution, since the capacity of DISCOP is about $83\%$ of the entropy. As for robustness, both are not robust to substitution. Even if the error probability of the channel is only $1\%$, none of their stegotexts can be correctly decoded, which is not unexpected since their constructions do not consider any possible modification. Compared to these steganography methods, our method provides a relatively low capacity, but has significant advantages in decoding rate and robustness. In particular, our decoding algorithm does not require the execution of the model, while the decoding algorithms of METEOR and DISCOP do not work without the model.

For watermark methods, we observe an explicit increase in perplexity from the results of Red-Green List. Since this method modifies the distribution of model's output, adversaries may have a noticeable probability of distinguishing between watermarked output and normal output. On the contrary, the Undetectable Watermark produces computationally indistinguishable stegotext and the perplexity and entropy are similar to the covertext. However, this method requires a large amount of text to embed the watermark. With the secure parameter $\lambda = 16$, this method still needs over $200$ tokens to embed the watermark. Compared to the Red-Green List, our method generates more qualified stegotext. Compared to the Undetectable Watermark, our method requires less tokens to embed a single bit. 

\section{Conclusion}

In this paper, we design a steganography framework for the asymmetric resource scenario. This steganography framework is based on the concepts of hypothesis testing, and its security, correctness, and robustness are rigorously proven. To the best of our knowledge, this work is the first practical and provably secure and robust steganography technique. We test our framework on some popular LLMs and report its performance in a practical scenario. The stegotext generated by our method shows equivalent quality to the normal output of the model. Moreover, our framework shows robustness against BSC with an error probability of about 7\%. This steganography framework expands the applicable range of steganography, and we hope it will be the foundation of a new branch of practical steganography. Our codes are available and a test website is open, details are shown in Appendix \ref{codes}.

\bibliography{custom,anthology}
\bibliographystyle{IEEEtran}
\begin{figure*}[ht]
    \centering
    \includegraphics[width=\linewidth]{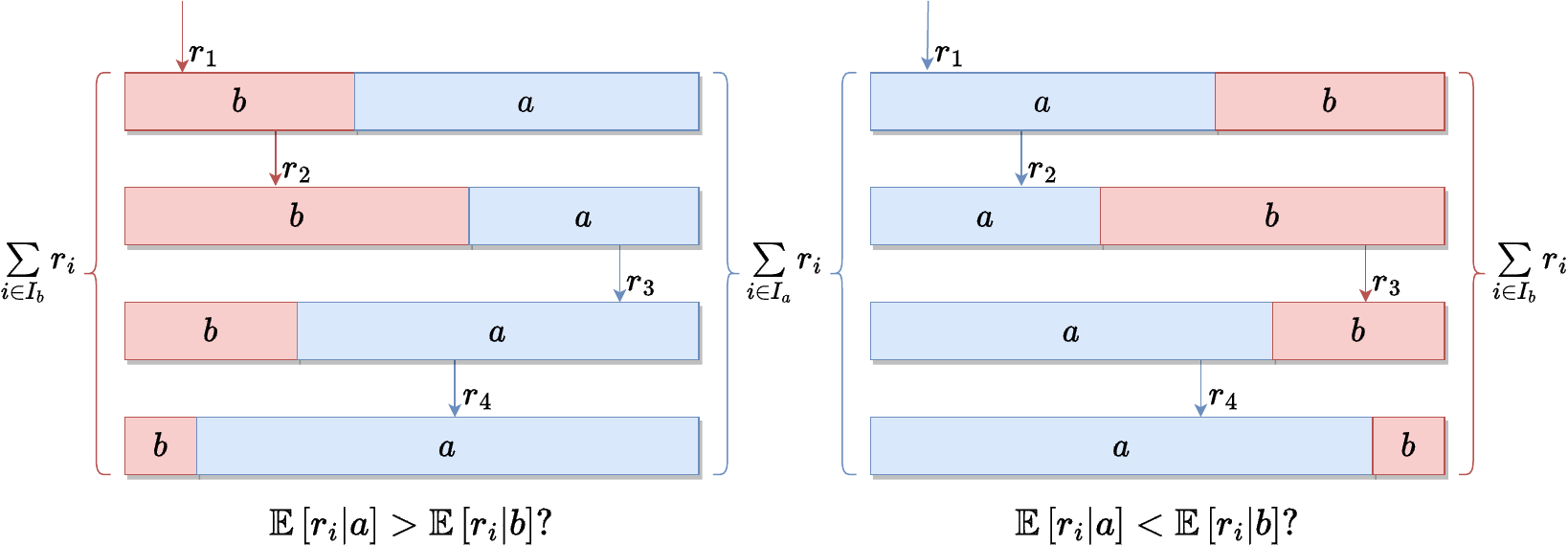}
    \caption{If we do not know the exact probability of symbol $a$ and $b$, we can compute the conditional expectation of the random numbers when receiving $a$ or $b$ to guess the permutation.}
    \label{fig:猜测.drawio.pdf}
\end{figure*}
\begin{figure*}[ht]
    \centering
    \includegraphics[width = \textwidth]{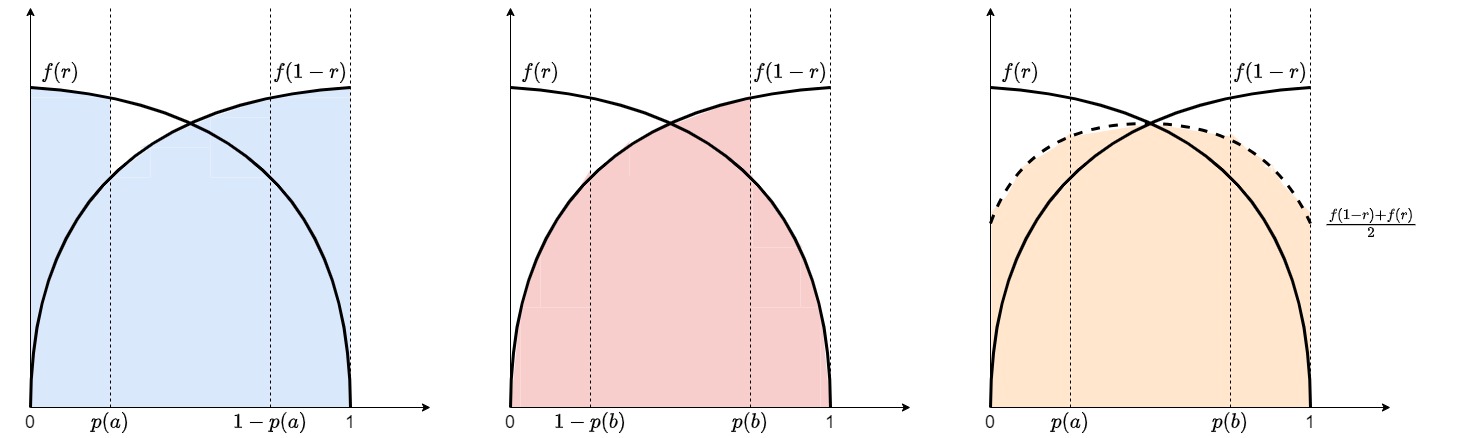}
    \caption{Computing the integral under $f(r)$ and $f(1-r)$. The left and middle figures show the expectation of $s((a,b),r)$ under permutation $(a,b)$ and $(b,a)$ respectively. The right figure shows the expectation of $s((a,b),r)$ when $r$ is independent from sampling.}
    \label{fig:积分.jpg}
\end{figure*}

\newpage

\appendices
\section{Intuition}\label{intuition}

Since the sum of all probabilities in the distribution is 1, we can visualize this as the interval $[0,1]$ divided into numerous smaller segments, each representing the probability of a corresponding symbol. Interestingly, different permutations of these segments within the $[0,1]$ interval can lead to notable variations.

We first consider the simplest distribution consisting of two symbols, $a$ and $b$, with probabilities $p(a)$ and $p(b)$, respectively, as illustrated in Fig. \ref{fig:顺序.drawio.pdf}. There are two possible permutations. In the first, symbol $a$ occupies the interval $[1 - p(a), 1]$ while symbol $b$ occupies the interval $[0, p(b)]$; we denote this permutation as ``$(b, a)$''. In the second, symbol $a$ occupies the interval $[0, p(a)]$ while symbol $b$ occupies the interval $[1 - p(b), 1]$; this permutation is denoted as ``$(a, b)$''. 

In each sampling procedure, a fresh random number will be drawn. If the random number $r$ falls into the interval that symbol $a$ occupies, symbol $a$ is sampled. Since the length of the interval that symbol $a$ occupies equals to the probability of symbol $a$, this sampling procedure does not change the original distribution, as the probability of output $a$ is $\mathbb{P}\left\{r < p(a)\right\} = p(a)$ and the probability of output $b$ is $\mathbb{P}\left\{r \geq p(a)\right\} = p(b)$. If the random number is drawn from a random source, the above derivation always holds. If the random number is drawn from a PRG with a secure parameter $\lambda$, we say that the sampling procedure is computationally indistinguishable from the random sampling.


Now assume that the encoder and the decoder can share the distribution $p(a) = 0.6$ and $p(b) = 0.4$. If encoder use the permutation $(b,a)$, when the symbol $b$ is sampled the decoder can know that the random number $r \in [0,0.4]$, and the decoder can check this by drawing a random number from the PRG using the same key that the encoder used. If the decoder find that exactly the random number $r \in [0,0.4]$, the decoder can conclude that the permutation is $(b,a)$. If the decoder find that the random number $r \not\in [0,0.4]$, he may think this is not a stegotext, or the stegotext has been modified during transmission. But, if the random number $r \in [0.4,0.6]$, the symbol $a$ will always be sampled regardless of the permutation, which makes the decoder cannot know the used permutation. So in this case, we say that the probability of the encoder to successfully let the decoder know the permutation is $0.8$, or we say that the probability of the encoder to successfully transmit a bit is $0.8$, since there are only 2 different permutations.

If the decoder cannot get access to the distribution, still he can guess that which permutation is used. Since the PRG and the key is shared, the decoder can always know the random number $r$ that the encoder uses in each step. If the symbol $b$ is sampled and random number $r$ is large, decoder will assume that the symbol $b$ occupies the right. If the symbol $b$ is sampled and random number $r$ is small, he will assume that the symbol $b$ occupies the left. As shown in Fig.\ref{fig:猜测.drawio.pdf}, the decoder can compare the conditional expectation $\mathbb{E}\left[r_i|a\right]$ and $\mathbb{E}\left[r_i|b\right]$ to find the used permutation.
We can let any of the permutation represent a bit $0$ or $1$, then we can do some steganography with these permutations and the decoder does not need to know the distribution.

\section{Compute the Optimal Function $f$}
According to the form of central inequalities, we aim at minimize the variance of $s|\mathcal{H}_\emptyset$ while maximize the gap between $\mathbb{E}\left[s|\mathcal{H}_\emptyset\right]$ and $\mathbb{E}\left[s|\mathcal{H}_0\right]$. The optimization objective is 
\begin{align}\label{opt}
    \min\limits_f \frac{\int_0^1 f^2(r) dr - \left(\int_0^1 f(r) dr \right)^2}{\left|\mathbb{E}_{p}\left[\int_0^{p} f(r) - f(1-r)dr\right]\right|},
\end{align}
 
Let $g(r) = f(r) - \int_0^1 f(r) dr$, then $\int_0^1 g(r) dr  = 0$ and 
\begin{align}
    &\int_0^1 g^2(r) dr \notag \\
    = & \int_0^1 \left(f(r) - \int_0^1 f(r) dr\right)^2 dr \notag \\
    = & \int_0^1 f^2(r) dr - 2 \int_0^1 f(r) dr\int_0^1 f(r) dr + \left(\int_0^1 f(r) dr\right)^2 \notag \\
    = & \int_0^1 f^2(r) dr - \left(\int_0^1 f(r) dr \right)^2.
\end{align}
So the optimization problem can be simplified to the following form
\begin{align}
        \min\limits_g \frac{\int_0^1 g^2(r) dr}{\left|\mathbb{E}_{p}\left[\int_0^{p} g(r) - g(1-r)dr\right]\right|}.
\end{align}
First we focus on the upper part. Here we need to expand the real functions $g$ and $G$ to complex functions. From the Cauchy-Schwarz Inequality, 
\begin{align}
    \left|\int_0^{1} g(p)\overline{G(p)} dp\right|^2 \leq \int_0^{1} \left|g(p)\right|^2 dp \int_0^{1} \left|G(p)\right|^2 dp.
\end{align}
The condition for equality is $\overline{G(p)} = \alpha g(p)$, where $\alpha$ is a constant. Since $\int_0^1 g(r)dr = 0 $, we let $g(p) = e^{i \pi p}$, $G(p) = \frac{1}{i \pi}e^{i \pi p}$,
left term of the above inequality can be simplified as
\begin{align}
    &\left|\int_0^{1} g(p)\overline{G(p)} dp\right| \notag \\
    = & \left|\int_0^{1} e^{i \pi p}\frac{1}{-i \pi}e^{-i \pi p} dp\right| 
    = \left|\int_0^{1} \frac{i}{\pi} dp\right| 
    = \frac{1}{\pi}.
\end{align}
And 
\begin{align}
    \int_0^{1} \left|G(p)\right|^2 dp = \frac{1}{\pi^2}.
\end{align}
So
\begin{align}
    \int_0^{1} \left|g(p)\right|^2 dp \geq \frac{\left|\int_0^{1} g(p)\overline{G(p)} dp\right|^2}{\int_0^{1} \left|G(p)\right|^2 dp} = 1
\end{align}
Then we focus on the bottom part.
\begin{align}
    & \mathbb{E}_{p}\left|\int_0^{p} g(r) - g(1-r)dr\right| \notag\\
    = & \left|\int_0^1 \text{pdf}(p) \left[\int_0^{p} g(r) - g(1-r)dr\right] dp\right| \notag \\
    = & \left|\int_0^1 \text{pdf}(p) \left[G(p) + G(1-p) - G(0) - G(1)\right] dp \right| \notag \\
    = & \left|2\int_0^1 \text{pdf}(p) G(p)dp - 2G(0) \right| 
    \notag \\ 
    \leq & 2\left|\frac{\int_0^1 G(p)\overline{G(p)}dp}{\int_0^1 \overline{G(p)}dp}\right| + 2\left|G(0)\right| \notag \\
    = & 2\left(\frac{1}{2} + \frac{1}{\pi}\right)
    = 1+ \frac{2}{\pi}.
\end{align}

According to the above derivation, the optimization target is at least
\begin{align}
    & \frac{\int_0^1 \left|g(r)\right|^2 dr}{\left|\mathbb{E}_{p}\left[\int_0^{p} g(r) - g(1-r)dr\right]\right|}
    \geq \frac{1}{\left(1 + \frac{2}{\pi} \right)}
    = \frac{\pi}{\pi + 2}.
\end{align}
And the maximum of capacity is 
\begin{align}
    n > & \frac{2\left(g(0)-g(1)\right)^2(-\ln(P_e))}{\left(2\int_0^1 G(p)\text{pdf}(p) dp - 2G(0)\right)^2} \notag \\
    \geq & \frac{8(-\ln(P_e))}{\left( 1+\frac{2}{\pi} \right)}\notag \\
    = & \frac{8\pi}{\pi+2}(-\ln(P_e))
\end{align}

\textbf{The above derivation is not formal or rigorous, and the result may not be correct.}
\begin{figure*}[ht]
\centering
    \includegraphics[width=1\linewidth]{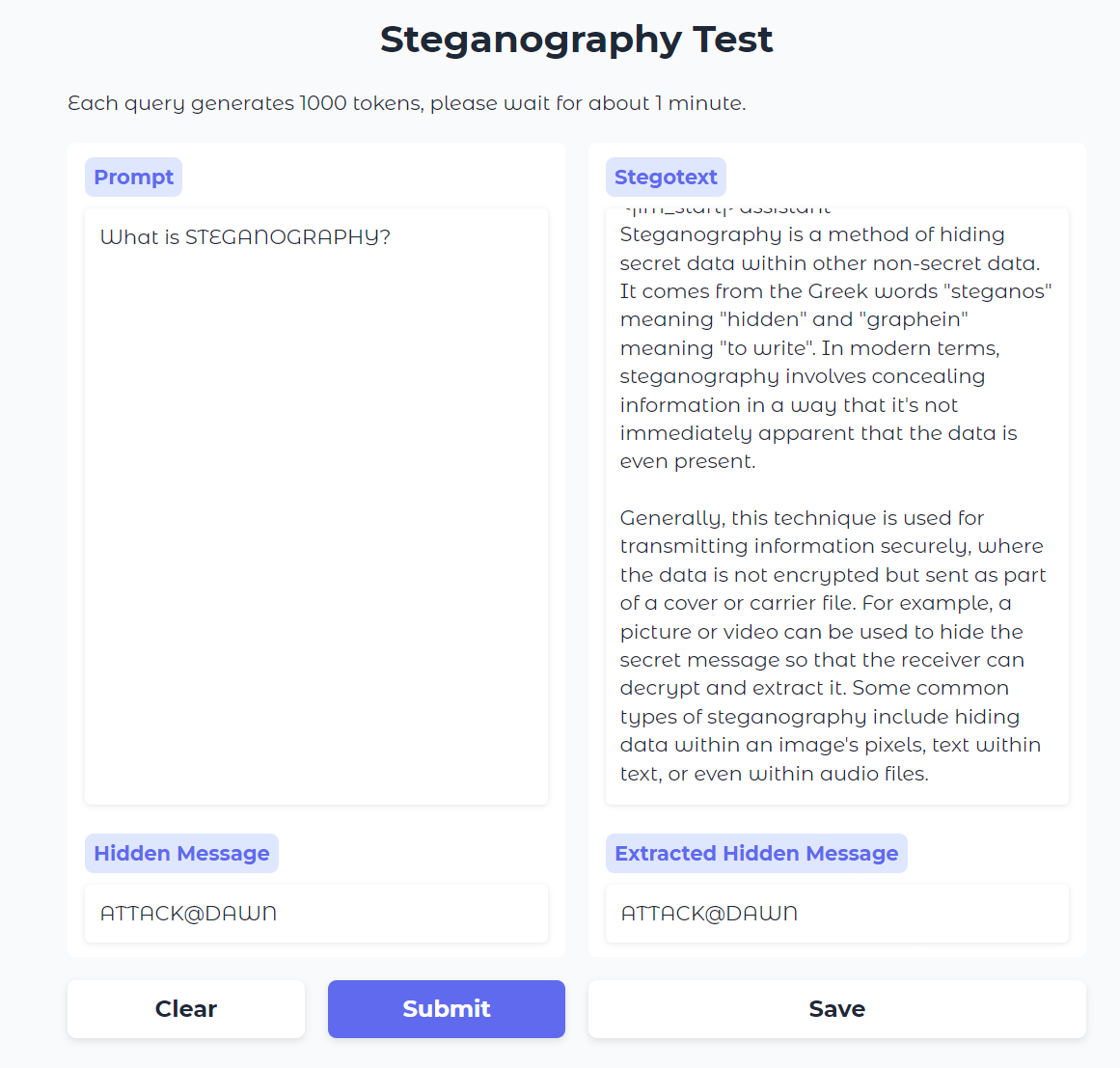}
    \caption{Our test website, constructed by Gradio.}
    \label{gradio}
\end{figure*}

\section{Codes \& Application}\label{codes}
Our codes are available at \url{https://anonymous.4open.science/r/Provably-robust-and-secure-steganography-0CBA}. We construct an online test website by Gradio as shown in Fig. \ref{gradio}. We plan to design a complex system for scenarios where multiple people are speaking simultaneously, aiming to extract the correct hidden messages from the mixed audio signals.

\end{document}